\documentclass[letterpaper]{IEEEtran}

\usepackage[numbers,sort&compress]{natbib}

\makeatletter

\usepackage{etex}

\usepackage{zref-savepos}

\newcounter{mnote}

\def\xmarginnote{%
  \xymarginnote{\hskip -\marginparsep \hskip -\marginparwidth}}

\def\ymarginnote{%
  \xymarginnote{\hskip\columnwidth \hskip\marginparsep}}

\long\def\xymarginnote#1#2{%
\vadjust{#1%
\smash{\hbox{{%
        \hsize\marginparwidth
        \@parboxrestore
        \@marginparreset
\footnotesize #2}}}}}

\def\mnoteson{%
\gdef\mnote##1{\refstepcounter{mnote}\label{##1}%
  \zsavepos{##1}%
  \ifnum20432158>\number\zposx{##1}%
  \xmarginnote{{\color{blue}\bf $\langle$\arabic{mnote}$\rangle$}}%
  \else
  \ymarginnote{{\color{blue}\bf $\langle$\arabic{mnote}$\rangle$}}%
  \fi%
}
  }
\gdef\mnotesoff{\gdef\mnote##1{}}
\mnoteson
\mnotesoff

\newcommand{\figref}[1]{Fig.~\ref{#1}}

\usepackage{framed}
\usepackage{subcaption}
\usepackage{comment}
\usepackage[svgnames,dvipsnames]{xcolor}
\usepackage[all]{xy}
\usepackage{tikz}
\usetikzlibrary{positioning,matrix,through,calc,arrows,fit,shapes,decorations.pathreplacing,decorations.markings,}

\tikzstyle{block} = [draw,fill=blue!20,minimum size=2em]

\tikzset{%
  prefix node name/.code={%
    \tikzset{%
      name/.code={\edef\tikz@fig@name{#1 ##1}}
    }%
  }%
}

\@ifpackagelater{tikz}{2013/12/01}{
\newcommand{\convexpath}[2]{
	[create hullcoords/.code={
		\global\edef\namelist{#1}
		\foreach [count=\counter] \nodename in \namelist {
			\global\edef\numberofnodes{\counter}
			\coordinate (hullcoord\counter) at (\nodename);
		}
		\coordinate (hullcoord0) at (hullcoord\numberofnodes);
		\pgfmathtruncatemacro\lastnumber{\numberofnodes+1}
		\coordinate (hullcoord\lastnumber) at (hullcoord1);
	}, create hullcoords ]
	($(hullcoord1)!#2!-90:(hullcoord0)$)
	\foreach [evaluate=\currentnode as \previousnode using \currentnode-1,
	evaluate=\currentnode as \nextnode using \currentnode+1] \currentnode in {1,...,\numberofnodes} {
		let \p1 = ($(hullcoord\currentnode) - (hullcoord\previousnode)$),
		\n1 = {atan2(\y1,\x1) + 90},
		\p2 = ($(hullcoord\nextnode) - (hullcoord\currentnode)$),
		\n2 = {atan2(\y2,\x2) + 90},
		\n{delta} = {Mod(\n2-\n1,360) - 360}
		in 
		{arc [start angle=\n1, delta angle=\n{delta}, radius=#2]}
		-- ($(hullcoord\nextnode)!#2!-90:(hullcoord\currentnode)$) 
	}
}
}{
\newcommand{\convexpath}[2]{
	[create hullcoords/.code={
		\global\edef\namelist{#1}
		\foreach [count=\counter] \nodename in \namelist {
			\global\edef\numberofnodes{\counter}
			\coordinate (hullcoord\counter) at (\nodename);
		}
		\coordinate (hullcoord0) at (hullcoord\numberofnodes);
		\pgfmathtruncatemacro\lastnumber{\numberofnodes+1}
		\coordinate (hullcoord\lastnumber) at (hullcoord1);
	}, create hullcoords ]
	($(hullcoord1)!#2!-90:(hullcoord0)$)
	\foreach [evaluate=\currentnode as \previousnode using \currentnode-1,
	evaluate=\currentnode as \nextnode using \currentnode+1] \currentnode in {1,...,\numberofnodes} {
		let \p1 = ($(hullcoord\currentnode) - (hullcoord\previousnode)$),
		\n1 = {atan2(\x1,\y1) + 90},
		\p2 = ($(hullcoord\nextnode) - (hullcoord\currentnode)$),
		\n2 = {atan2(\x2,\y2) + 90},
		\n{delta} = {Mod(\n2-\n1,360) - 360}
		in 
		{arc [start angle=\n1, delta angle=\n{delta}, radius=#2]}
		-- ($(hullcoord\nextnode)!#2!-90:(hullcoord\currentnode)$) 
	}
}
}

\usepackage{qsymbols,amssymb,mathrsfs}
\usepackage{amsmath}
\usepackage[standard,thmmarks]{ntheorem}
\theoremstyle{plain}
\theoremsymbol{\ensuremath{_\vartriangleleft}}
\theorembodyfont{\itshape}
\theoremheaderfont{\normalfont\bfseries}
\theoremseparator{}

\theoremstyle{nonumberplain}
\theoremheaderfont{\scshape}
\theorembodyfont{\normalfont}
\theoremsymbol{\ensuremath{_\blacktriangleleft}}

\theoremnumbering{arabic}
\theoremstyle{plain}
\usepackage{latexsym}
\theoremsymbol{\ensuremath{_\Box}}
\theorembodyfont{\itshape}
\theoremheaderfont{\normalfont\bfseries}
\theoremseparator{}

\theorembodyfont{\upshape}
\theoremprework{\bigskip\hrule}
\theorempostwork{\hrule\bigskip}

\usepackage[overload]{empheq} 

\let\iftwocolumn\if@twocolumn
\g@addto@macro\@twocolumntrue{\let\iftwocolumn\if@twocolumn}
\g@addto@macro\@twocolumnfalse{\let\iftwocolumn\if@twocolumn}

\mathtoolsset{showonlyrefs=false,showmanualtags}
\let\underbrace\LaTeXunderbrace 
\let\overbrace\LaTeXoverbrace
\renewcommand{\eqref}[1]{\textup{(\refeq{#1})}} 
\newtagform{brackets}[]{(}{)}   
\usetagform{brackets}

\usepackage[Smaller]{cancel}

\PassOptionsToPackage{breaklinks,letterpaper,hyperindex=true,backref=false,bookmarksnumbered,bookmarksopen,linktocpage,colorlinks,linkcolor=BrickRed,citecolor=OliveGreen,urlcolor=Blue,pdfstartview=FitH}{hyperref}
\usepackage{hyperref}

\usepackage{graphicx,psfrag}
\graphicspath{{figure/}{image/}} 

\usepackage{diagbox} 

\usepackage{algorithm2e}
\usepackage{listings} 
\lstdefinelanguage{Maple}{
  morekeywords={proc,module,end, for,from,to,by,while,in,do,od
    ,if,elif,else,then,fi ,use,try,catch,finally}, sensitive,
  morecomment=[l]\#,
  morestring=[b]",morestring=[b]`}[keywords,comments,strings]
\DeclareMathAlphabet{\mathpzc}{OT1}{pzc}{m}{it}
\usepackage{upgreek} 
\usepackage{dsfont}  

\def\multi@nostar#1#2{%
  \expandafter\def\csname multi#1\endcsname##1{%
    \if ##1.\let\next=\relax \else
    \def\next{\csname multi#1\endcsname}     
    \expandafter\newcommand\csname #1##1\endcsname{#2}
    \fi\next}}

\def\multi@star#1#2{%
  \expandafter\def\csname #1\endcsname##1{#2}
  \multi@nostar{#1}{#2}
}

\newcommand{\multi}{%
  \@ifstar \multi@star \multi@nostar}

\multi*{rm}{\mathrm{#1}}
\multi*{mc}{\mathcal{#1}}
\multi*{op}{\mathop {\operator@font #1}}
\multi*{ds}{\mathds{#1}}
\multi*{set}{\mathcal{#1}}
\multi*{rsfs}{\mathscr{#1}}
\multi*{pz}{\mathpzc{#1}}
\multi*{M}{\boldsymbol{#1}}
\multi*{R}{\mathsf{#1}}
\multi*{RM}{\M{\R{#1}}}
\multi*{bb}{\mathbb{#1}}
\multi*{td}{\tilde{#1}}
\multi*{tR}{\tilde{\mathsf{#1}}}
\multi*{trM}{\tilde{\M{\R{#1}}}}
\multi*{tset}{\tilde{\mathcal{#1}}}
\multi*{tM}{\tilde{\M{#1}}}
\multi*{baM}{\bar{\M{#1}}}
\multi*{ol}{\overline{#1}}

\multirm  ABCDEFGHIJKLMNOPQRSTUVWXYZabcdefghijklmnopqrstuvwxyz.
\multiol  ABCDEFGHIJKLMNOPQRSTUVWXYZabcdefghijklmnopqrstuvwxyz.
\multitR   ABCDEFGHIJKLMNOPQRSTUVWXYZabcdefghijklmnopqrstuvwxyz.
\multitd   ABCDEFGHIJKLMNOPQRSTUVWXYZabcdefghijklmnopqrstuvwxyz.
\multitset ABCDEFGHIJKLMNOPQRSTUVWXYZabcdefghijklmnopqrstuvwxyz.
\multitM   ABCDEFGHIJKLMNOPQRSTUVWXYZabcdefghijklmnopqrstuvwxyz.
\multibaM   ABCDEFGHIJKLMNOPQRSTUVWXYZabcdefghijklmnopqrstuvwxyz.
\multitrM   ABCDEFGHIJKLMNOPQRSTUVWXYZabcdefghijklmnopqrstuvwxyz.
\multimc   ABCDEFGHIJKLMNOPQRSTUVWXYZabcdefghijklmnopqrstuvwxyz.
\multiop   ABCDEFGHIJKLMNOPQRSTUVWXYZabcdefghijklmnopqrstuvwxyz.
\multids   ABCDEFGHIJKLMNOPQRSTUVWXYZabcdefghijklmnopqrstuvwxyz.
\multiset  ABCDEFGHIJKLMNOPQRSTUVWXYZabcdefghijklmnopqrstuvwxyz.
\multirsfs ABCDEFGHIJKLMNOPQRSTUVWXYZabcdefghijklmnopqrstuvwxyz.
\multipz   ABCDEFGHIJKLMNOPQRSTUVWXYZabcdefghijklmnopqrstuvwxyz.
\multiM    ABCDEFGHIJKLMNOPQRSTUVWXYZabcdefghijklmnopqrstuvwxyz.
\multiR    ABCDEFGHIJKL NO QR TUVWXYZabcd fghijklmnopqrstuvwxyz.
\multibb   ABCDEFGHIJKLMNOPQRSTUVWXYZabcdefghijklmnopqrstuvwxyz.
\multiRM   ABCDEFGHIJKLMNOPQRSTUVWXYZabcdefghijklmnopqrstuvwxyz.

\newcommand{\dotleq}{\buildrel \textstyle  .\over {\smash{\lower
      .2ex\hbox{\ensuremath\leqslant}}\vphantom{=}}}
\newcommand{\dotgeq}{\buildrel \textstyle  .\over {\smash{\lower
      .2ex\hbox{\ensuremath\geqslant}}\vphantom{=}}}

\newcommand{\bM}{\begin{bmatrix}}
\newcommand{\eM}{\end{bmatrix}}
\newcommand{\bSM}{\left[\begin{smallmatrix}}
\newcommand{\eSM}{\end{smallmatrix}\right]}
\renewcommand*\env@matrix[1][*\c@MaxMatrixCols c]{%
  \hskip -\arraycolsep
  \let\@ifnextchar\new@ifnextchar
  \array{#1}}

\newqsymbol{`N}{\mathbb{N}}
\newqsymbol{`R}{\mathbb{R}}
\newqsymbol{`P}{\mathbb{P}}
\newqsymbol{`Z}{\mathbb{Z}}

\newqsymbol{`|}{\mid}
\newqsymbol{`8}{\infty}
\newqsymbol{`1}{\left}
\newqsymbol{`2}{\right}
\newqsymbol{`6}{\partial}
\newqsymbol{`0}{\emptyset}
\newqsymbol{`-}{\leftrightarrow}
\newqsymbol{`<}{\langle}
\newqsymbol{`>}{\rangle}

\DeclarePairedDelimiter\abs{\lvert}{\rvert}

\DeclarePairedDelimiter\Set{\{}{\}}
\newcommand{\imod}[1]{\allowbreak\mkern10mu({\operator@font mod}\,\,#1)}

\newcommand{\threecols}[3]{
\hbox to \textwidth{%
      \normalfont\rlap{\parbox[b]{\textwidth}{\raggedright#1\strut}}%
        \hss\parbox[b]{\textwidth}{\centering#2\strut}\hss
        \llap{\parbox[b]{\textwidth}{\raggedleft#3\strut}}%
    }
}

\newcommand{\reason}[2][\relax]{
  \ifthenelse{\equal{#1}{\relax}}{
    \left(\text{#2}\right)
  }{
    \left(\parbox{#1}{\raggedright #2}\right)
  }
}

\newcommand{\utag}[2]{\mathop{#2}\limits^{\text{(#1)}}}
\newcommand{\uref}[1]{(#1)}

\let\SavedDoubleVert\relax
\let\protect\relax
{\catcode`\|=\active
  \xdef\extendvert{\protect\expandafter\noexpand\csname extendvert \endcsname}
  \expandafter\gdef\csname extendvert \endcsname#1{\mskip-5mu \left.%
      \ifx\SavedDoubleVert\relax \let\SavedDoubleVert\|\fi
     \:{\let\|\SetDoubleVert
       \mathcode`\|32768\let|\SetVert
     #1}\:\right.\mskip-5mu}
}
\def\SetVert{\@ifnextchar|{\|\@gobble}
    {\egroup\;\mid@vertical\;\bgroup}}
\def\SetDoubleVert{\egroup\;\mid@dblvertical\;\bgroup}

\begingroup
 \edef\@tempa{\meaning\middle}
 \edef\@tempb{\string\middle}
\expandafter \endgroup \ifx\@tempa\@tempb
 \def\mid@vertical{\middle|}
 \def\mid@dblvertical{\middle\SavedDoubleVert}
\else
 \def\mid@vertical{\mskip1mu\vrule\mskip1mu}
 \def\mid@dblvertical{\mskip1mu\vrule\mskip2.5mu\vrule\mskip1mu}
\fi

\makeatother

\usepackage{fouridx}
\usepackage{framed}
\usetikzlibrary{positioning,matrix}

\usepackage{paralist}
\usepackage{enumerate}

\usepackage[normalem]{ulem}

\numberwithin{equation}{section}
\makeatletter
\@addtoreset{equation}{section}
\renewcommand{\theequation}{\arabic{section}.\arabic{equation}}
\@addtoreset{Theorem}{section}
\renewcommand{\theTheorem}{\arabic{section}.\arabic{Theorem}}
\@addtoreset{Lemma}{section}
\renewcommand{\theLemma}{\arabic{section}.\arabic{Lemma}}
\@addtoreset{Corollary}{section}
\renewcommand{\theCorollary}{\arabic{section}.\arabic{Corollary}}
\@addtoreset{Example}{section}
\renewcommand{\theExample}{\arabic{section}.\arabic{Example}}
\@addtoreset{Remark}{section}
\renewcommand{\theRemark}{\arabic{section}.\arabic{Remark}}
\@addtoreset{Proposition}{section}
\renewcommand{\theProposition}{\arabic{section}.\arabic{Proposition}}
\@addtoreset{Definition}{section}
\renewcommand{\theDefinition}{\arabic{section}.\arabic{Definition}}
\@addtoreset{Claim}{section}

\@addtoreset{Subclaim}{Theorem}
\renewcommand{\theSubclaim}{\theTheorem\Alph{Subclaim}}
\makeatother

\newenvironment{lbox}{
  \setlength{\FrameSep}{1.5mm}
  \setlength{\FrameRule}{0mm}
  \MakeFramed {\FrameRestore}}%
{\endMakeFramed}

\newenvironment{ybox}{
	\setlength{\FrameSep}{1.5mm}
	\setlength{\FrameRule}{0mm}
  \MakeFramed {\FrameRestore}}%
{\endMakeFramed}

\newenvironment{gbox}{
	\setlength{\FrameSep}{1.5mm}
\setlength{\FrameRule}{0mm}
  \MakeFramed {\FrameRestore}}%
{\endMakeFramed}

{\endMakeFramed}

 {\endMakeFramed}

\usepackage{enumitem}

\usepackage{etoolbox}

\let\theparentequation\theequation
\patchcmd{\theparentequation}{equation}{parentequation}{}{}

\renewenvironment{subequations}[1][]{
	\refstepcounter{equation}%
	\setcounter{parentequation}{\value{equation}}
	\setcounter{equation}{0}
	\def\theequation{\theparentequation\alph{equation}}%
	\let\parentlabel\label
	\ifx\\#1\\\relax\else\label{#1}\fi
	\ignorespaces
}{%
	\setcounter{equation}{\value{parentequation}}
	\ignorespacesafterend
}

\newcommand*{\nextParentEquation}[1][]{
	\refstepcounter{parentequation}
	\setcounter{equation}{0}
	\ifx\\#1\\\relax\else\parentlabel{#1}\fi
}

\newcommand{\RCO}{R_{\op{CO}}}

\newcommand{\RS}{R_{\op{S}}}
\newcommand{\CS}{C_{\op{S}}}

\usepackage[outline]{contour}
\contourlength{1.2pt}

\tikzstyle{dot}=[circle,draw=gray!80,fill=gray!20,thick,inner 
sep=2pt,minimum size=11pt]
\tikzstyle{edge}=[-,draw=blue]
\tikzstyle{point}=[draw,circle,minimum size=.2em,inner sep=0, outer sep=.2em]
\tikzstyle{hyperedge}=[blue,fill,fill opacity=0.1]

\title{Upper Bounds via Lamination on the Constrained Secrecy Capacity of Hypergraphical Sources}
\author{Chung Chan, Manuj Mukherjee, Navin Kashyap and Qiaoqiao Zhou
	\thanks{Parts of this work were presented at the 2017 IEEE International Symposium on Information Theory (ISIT 2017), Aachen, Germany~\cite{chan17isit}.}
        \thanks{C.\ Chan (corresponding author, email: chung.chan@cityu.edu.hk) is with the Department of Computer Science, City University of Hong Kong. His work was supported by a grant from the University Grants Committee of the Hong Kong Special Administrative Region, China (Project No. 14200714).}
	\thanks{Q.\ Zhou is with the Department of Information Engineering, the
          Chinese University of Hong Kong.}
        \thanks{N.\ Kashyap (nkashyap@iisc.ac.in) is with the Department of Electrical Communication Engineering, Indian Institute of Science, Bangalore 560012. His work was supported in part by a Swarnajayanti Fellowship awarded to N.\ Kashyap by the Department of Science \& Technology, Government of India.}
        \thanks{M.\ Mukherjee (manuj.mukherjee@telecom-paristech.fr) is with the Digital Communications Group, Communications and Electronics (Comelec) Department, Telecom Paristech, Paris, France.}}

\begin{document}
	
\maketitle
	
\begin{abstract}
Hypergraphical sources are a natural class of sources for secret key generation, within which different subsets of terminals sharing secrets are allowed to discuss publicly in order to agree upon a global secret key. While their secrecy capacity, i.e., the maximum rate of a secret key that can be agreed upon by the entire set of terminals, is well-understood, what remains open is the maximum rate of a secret key that can be generated when there is a restriction on the overall rate of public discussion allowed. In this work, we obtain a family of explicitly computable upper bounds on the number of bits of  secret key that can be generated per bit of public discussion. These upper bounds are derived using a lamination technique based on the submodularity of the entropy function. In particular, a specific instance of these upper bounds, called the edge-partition bound, is shown to be tight for the pairwise independent network model, a special case of the hypergraphical source. The secret key generation scheme achieving this upper bound is the tree-packing protocol of Nitinawarat et al., thereby resolving in the affirmative the discussion rate optimality of the tree packing protocol.
\end{abstract} 


\section{Introduction}

The problem of secret key generation between a pair of terminals was independently proposed by Maurer \cite{maurer93}, and Ahlswede and Csisz\'{a}r \cite{ahlswede93}. The pair of terminals are allowed to \emph{interactively} discuss in public over a noiseless broadcast channel in order to agree upon a secret key, which is to be secured from a \emph{passive} eavesdropper who monitors the communication sent over the public channel. The problem was later extended to the case of multiple terminals observing correlated sources by Csisz\'{a}r and Narayan \cite{csiszar04}. The quantity of interest in all of these works was the \emph{secrecy capacity}, i.e., the maximum rate of a secret key that can be agreed upon by all the terminals. However, these works treated communication as a free resource, an assumption which does not hold in practical scenarios. In fact, Csisz\'{a}r and Narayan \cite{csiszar04} showed using some examples that their \emph{communication for omniscience} strategy used to achieve secrecy capacity may require strictly more communication than needed.

The first work to consider the effects of \emph{rate-limited} communication on the secret key generation problem is due to Csisz\'{a}r and Narayan \cite{csiszar00}. The authors derived a complete characterization of the key-rate versus communication-rate tradeoff for the two-terminal scenario where only one-way discussion is allowed. Later, Tyagi \cite{tyagi13} looked at the problem of characterizing the \emph{communication complexity}, i.e., the minimum rate of interactive communication needed to achieve the secrecy capacity, for two-terminal sources. He obtained a multi-letter expression for the communication complexity using the \emph{interactive common information}, a quantity related to the \emph{Wyner common information} \cite{wyner75}, of the two sources. Tyagi left open the question of characterizing the key-rate versus interactive communication-rate tradeoff for two-terminal sources. This question was later partially addressed by Liu et al. \cite{LCV16}, who gave a complete characterization of the key-rate versus communication-rate tradeoff for a fixed number of communication rounds, using the ideas of interactive source coding developed by Kaspi \cite{Kaspi}. A complete and computable characterization of the key-rate versus communication-rate tradeoff for two-terminal sources with no restriction on the number of communication rounds is still open. Liu et al. also studied the quantity $\frac{\CS(R)}{R}$, where $\CS(R)$ is the maximum rate of secret-key with the rate of public discussion restricted to $R$. Using the notion of a \emph{symmetric strong data processing constant}, the authors derived the behaviour of the ratio $\frac{\CS(R)}{R}$ in two regimes, when $R$ goes to $0$, and when $\CS(R)$ is close to the secrecy capacity. While the above mentioned works all involve sources with finite alphabets, the case with Gaussian sources have also been considered. The characterization of the key-rate versus communication-rate tradeoff for two-terminal scalar and vector Gaussian sources has been carried out by Watanabe and Oohama in \cite{WO10} and \cite{WO10v}.

While the problem for two-terminal sources has received a fair bit of attention, literature on the multiterminal scenario is scant. Attempts have been made to obtain bounds on the communication complexity for multiterminal sources in \cite{MKS16} and \cite{mukherjee16}. In \cite{MKS16}, a lower bound on communication complexity has been derived by extending Tyagi's definition of interactive common information to a multiterminal scenario. Upper bounds on communication complexity have been developed in \cite{mukherjee16} using the idea of \emph{decremental secret key agreement} \cite{chan16isit}. Another direction of investigation has been characterizing multiterminal sources, for which the communication for omniscience protocol of Csisz\'{a}r and Narayan is communication-rate-optimal for achieving secrecy capacity. A sufficient condition to check the optimality of the communication for omniscience was derived in \cite{chan16itw}, and extensions of this result to sources involving helpers, untrusted terminals, and silent terminals was carried out in \cite{chan17oo}. While the above mentioned works look at the near secrecy capacity regime, the zero communication-rate regime has been investigated in \cite{chan18isit} for the special case of \emph{finite linear sources}. 

In this paper, we study the key-rate versus communication-rate tradeoff for multiterminal sources. At the outset we must mention that a study of general multiterminal sources is difficult, and hence, we shall restrict our attention to a specific class of sources, namely, the \emph{hypergraphical source} \cite{chan10md}. To explain our choice, consider the following natural scenario for secret key agreement. Certain subsets of terminals already possess secrets shared locally among themselves, and the terminals must agree upon a globally shared secret through public discussion. Let us ask this simple question: \emph{How many bits of globally shared secret can be generated using locally shared secrets?} The scenario described can be viewed as a hypergraphical source. The hypergraphical source consists of certain subsets of terminals observing i.i.d. sequences of random variables, which can be thought of as the local secrets. Therefore, the answer to the question posed earlier is simply the secrecy capacity of the hypergraphical source. Hypergraphical sources also appeared in the \emph{coded cooperative data exchange} (CCDE) problem \cite{ESS10}, \cite{courtade16}. 

The main contribution of this work is obtaining upper bounds on the ratio $\frac{\CS(R)}{R}$ for hypergraphical sources. Unlike earlier works on the two-terminal scenario, our results are not restricted to any particular regime and hold for every possible communication rate $R$. The upper bounds on $\frac{\CS(R)}{R}$ studied here are based on the fact that entropy is a \emph{submodular} set function~\cite{fujishige78}. Along with the specialized structure of the hypergraphical source, the submodularity of entropy enables us to define a `lamination' procedure which serves as the key ingredient to derive our bounds. The lamination procedure we use essentially boils down to minimizing a weighted sum of submodular functions using \emph{Edmonds' Greedy Algorithm} \cite[Theorem~44.3]{schrijver02}. In particular, we obtain three different upper bounds to $\frac{\CS(R)}{R}$ by laminating three different sums of entropies. The first of these bounds, which we shall call the \emph{edge-partition (EP) bound}, gives us an exact characterization of the key-rate versus communication-rate tradeoff for the so-called \emph{pairwise independent network} (PIN) model \cite{nitinawarat-ye10}, \cite{nitinawarat10}, which is a special case of the hypergraphical source. The tightness of the EP bound for the PIN model is shown using the \emph{tree-packing protocol} of Nitinawarat and Narayan \cite{nitinawarat10}. We would like to highlight that this is the first result which completely characterizes the key-rate versus communication-rate tradeoff for a large class of sources, without any restriction on the number of rounds of interactive communication. Also, the tradeoff does not involve any auxiliary random variables, and in fact, it can be expressed simply in terms of the size of the network. While the EP bound gives tight results for the PIN model, we show using an example that it can be loose for certain hypergraphical sources. To circumvent this issue, we derive our second upper bound, which we call the \emph{vertex-packing (VP) bound}. Although, the VP bound is tight for certain examples where the EP bound is loose, there are examples where the VP bound is loose but the EP bound is tight as well. To get the best of both the VP and EP bounds, we generalize them to obtain a third bound which we simply call the \emph{lamination bound}.

The paper is organized as follows. Section~\ref{sec:problem} introduces the hypergraphical source and states the necessary definitions. Section~\ref{sec:prelim} describes the tree-packing protocol for the PIN model. The main results of the paper, which include the EP bound, the VP bound, and the lamination bound are presented in Section~\ref{sec:main}. The contributions made by the paper, as well as future directions of research are summarized in Section~\ref{sec:conclusion}. The proofs of some of the technical results appear in the appendices.

\section{Problem Formulation}
\label{sec:problem}

We consider the basic source model for multiterminal secret key
agreement in \cite{csiszar04} but with no helpers and wiretapper's
side information. It involves a finite set 
$V$ of at least $2$ users. Without loss of generality, we can set $V$
to be $[m]:=\{1,2,\ldots,m\}$
with $m\geq 2$. The users have access to a private (discrete
memoryless multiple) source, which is denoted by the random vector
\begin{align*}
  \RZ_V & := (\RZ_i\mid i\in V).
\end{align*}
We assume that the random vector takes values from a finite set denoted
by
\begin{align*}
  Z_V & := \prod\nolimits_{i\in V} Z_i.
\end{align*}
Note that we use capital letters in sans serif font for random variables
and the corresponding capital letters in the usual math italic font
for the alphabet sets if there is no ambiguity. $P_{\RZ_V}$ denotes
the joint distribution of the $\RZ_i$'s.

The users want to agree on a secret key via public discussion. As in
\cite{csiszar04}, the protocol is divided into the following phases:
\begin{lbox}
  \noindent\underline{Private observation:} Each user $i\in V$ observes an $n$-sequence 
  \begin{align*}
    \RZ_i^n:=(\RZ_{it}\mid t\in[n])=(\RZ_{i1},\RZ_{i2},\ldots,\RZ_{in})
  \end{align*}
  i.i.d. generated from the source $\RZ_i$ for some block length $n$.
  
  \noindent\underline{Private randomization:} Each user $i\in V$ generates a random variable $\RU_i$ independent of the private source, i.e.,
  \begin{align}
    \label{eq:U}
    H(\RU_V|\RZ_V)=\sum_{i\in V}H(\RU_i).
  \end{align}
  For convenience, we denote the entire private observation of user $i\in V$ as 
  \begin{align}
    \tRZ_i:=(\RU_i,\RZ_i^n).\label{eq:tRZ}
  \end{align}
  
  \noindent\underline{Public discussion:}  Using a public
  authenticated noiseless channel, each user $i\in V$ broadcasts a
  message $\RF_{it}$ in round $t\in [\ell]$ for some positive integer $\ell$ number of rounds. The message is chosen as
  \begin{subequations}
    \label{eq:discussion}
    \begin{align}
      \RF_{it}&:=f_{it}(\tRZ_i,\tRF_{it}), \label{eq:Fit}      
    \end{align}
    which is a function of the accumulated observations of user~$i$,
    namely, his private observation $\tRZ_i$ defined in
    \eqref{eq:tRZ}, and the previous discussion
    \begin{alignat}{2}
      \tRF_{it} &:=(\RF_{[i-1]t},\RF_V^{t-1}),\label{eq:tFit}
    \end{alignat}
    where the first part $\RF_{[i-1]t}$ consists of the previous
    messages broadcast in the same round, and the second part
    $\RF_V^{t-1}$ denotes the messages broadcast in the previous
    rounds. Without loss of generality, we have assumed that the
    \emph{interactive discussion} is conducted in the ascending order
    of user indices. For convenience, we also write
    \begin{align}
      \RF_i & := \RF_{i[\ell]}=(\RF_{it}\mid t\in[\ell])\label{eq:Fi}\\
      \RF & := \RF_V = (\RF_i\mid i\in V)\label{eq:F}
    \end{align}
    to denote, respectively, the aggregate message from user $i\in V$ and the aggregation of the messages from all users.
  \end{subequations}
  \noindent\underline{Key generation:} A random variable $\RK$, called the secret key, is required to satisfy the \emph{recoverability} constraint that
  \begin{equation}
    \lim_{n\to\infty}\text{Pr}(\exists i\in V, \RK\neq\theta_i(\tRZ_i,\RF))=0, \label{eq:recover}
  \end{equation}
  for some function $\theta_i$, and the \emph{secrecy} constraint that
  \begin{equation}
    \lim_{n\to\infty}\frac{1}{n}`1[\log\abs{K}-H(\RK|\RF)`2]=0, \label{eq:secrecy}
  \end{equation}
  where $K$ denotes the finite alphabet set of possible key values.
\end{lbox}

It is desirable to have a large secret key rate $\frac1n\log\abs{K}$ but a small public discussion rate $\frac1n\log\abs{F}$. Our goal is
to characterize \emph{the optimal tradeoff between the secret key rate and
the total discussion rate}:

\begin{Definition}
  \label{def:CSR}
  The (total-discussion-rate-)\emph{constrained secrecy capacity} is defined for $R\geq 0$ as 
  \begin{equation}
    \CS(R):=\sup \liminf_{n\to\infty}\frac{1}{n}\log\abs{K} \label{eq:CSR}
  \end{equation}
  where the supremum is taken over all possible sequences of $(\RU_V , \RF, \RK)$ that satisfies the sum rate constraint on the public discussion
  \begin{equation}
    R \geq\limsup_{n\to\infty}\frac{1}{n}\log\abs{F}, \label{eq:rate}
  \end{equation}
  in addition to \eqref{eq:recover} and \eqref{eq:secrecy}. 
\end{Definition}

The curve $\CS(R)$ for $R\geq 0$ exists and is well-behaved with the following basic properties.
\begin{Proposition}\label{prop:CS(R)}
	$\CS(R)$ is continuous, non-decreasing and concave for $R\geq 0$.
\end{Proposition}
\begin{Proof}
	Continuity is because the $\liminf$ and $\limsup$ in \eqref{eq:rate} always exist, since $\CS(R)$ is bounded within $[0,H(\RZ_V)]$. The monotonicity is obvious, and concavity follows from the usual time sharing argument.
\end{Proof}

As motivated in the introduction, we
will restrict to the hypergraphical source model defined below:
\begin{Definition}[Definition~2.4 of \cite{chan10md}]\label{def:hyp}
  $\RZ_V$ is a \emph{hypergraphical source} with respect to  a hypergraph
  $(V,E,`x)$ with edge set $E$ and edge function $`x:
  E\to2^V`/\{\emptyset\}$ iff, for some mutually independent
  (hyper)edge (random) variables $\RX_e$ for $e\in E$, we can write
  \begin{equation}
    \RZ_i=(\RX_e\mid  e\in E, i\in`x(e)), \text{ for } i\in V, \label{eq:Xe}
  \end{equation}
  The \emph{weight function} $c:2^V`/\{\emptyset\}\to\mathbb{R}$ of a hypergraphical source is defined as
  \begin{subequations}
    \label{eq:c}
    \begin{align}
      c(B)&:=\sum_{e\in E: `x(e)=B}H(\RX_e) \kern.5em \text{with support}\kern-.5em\\
      \kern-.5em \op{supp}(c)&:=\Set*{B\in 2^V`/\{\emptyset\} \mid c(B)>0}
    \end{align}
  \end{subequations}
\end{Definition}

For convenience, we further make some mild assumptions on the hypergraphical
sources we will consider:
\begin{enumerate}
\item Every edge variable is non-trivial, i.e., $H(\RX_e)>0$ for all $e\in E$.
\item There exists at least one edge variable, i.e., $\abs{E}>0$.
\item No edge covers the entire set, i.e., $`x(e)\subsetneq V$ for all
  $e\in E$.
\end{enumerate}
The first assumption is without loss of generality, the second is to
avoid triviality. The last assumption is for simplicity.\footnote{It is possible to extend the results
of this work to allow for edges covering
the entire set: The corresponding edge variables can be used directly
as the secret key after simple
source compression, without any additional public discussion.} Note
also that the two-user case is also trivial, with $\CS(R)=0$, and so
we focus on the case $\abs{V}>2$.

An example of a hypergraphical source is as follows.
\begin{Example}
  \begin{figure}
    \centering 
    \def\u{1em}
    \begin{tikzpicture}[x=.6em,y=.6em,>=latex]
      \scriptsize
      \path (90:5*\u) node (1) [dot] {1};
      \path (-30:5*\u) node (2) [dot] {2};
      \path (210:5*\u) node (3) [dot] {3};
      \path (1) +(-9*\u,0*\u) node [dot] (4) {4};
      \path (1) +(9*\u,0*\u) node [dot] (5) {5};

      \draw[hyperedge] \convexpath{1,2,3}{1.4*\u};
      \node at ($0.5*(hullcoord2)+0.5*(hullcoord3)$) [yshift=-2*\u] {$\rm{a}$};
      \draw[hyperedge] \convexpath{1,3,4}{1.8*\u};
      \node at ($0.5*(hullcoord2)+0.5*(hullcoord3)$) [xshift=-2*\u,yshift=-2*\u] {$\rm{b}$};
      \draw[hyperedge] \convexpath{1,2,4}{2.2*\u};
      \node at ($0.5*(hullcoord0)+0.5*(hullcoord1)$) [xshift=0*\u,yshift=3*\u] {$\rm{c}$};
      \draw[hyperedge] \convexpath{1,5}{1*\u};
      \node at ($0.5*(hullcoord1)+0.5*(hullcoord2)$) [xshift=0*\u,yshift=2*\u] {$\rm{d}$};

      \foreach \x in {1,2,...,5}
      \node at (\x) [dot] {\x};

    \end{tikzpicture}
    \caption{The hypergraph corresponding to the hypergraphical source defined in \eqref{eq:receptacle}. Each edge $e$ corresponds to an independent edge variable $\RX_e$ in the private observation $\RZ_i$ associated with each incident node (user) $i$.}
    \label{fig:receptacle}
  \end{figure}
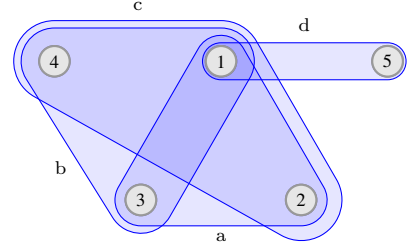
  Let $\RX_{\rm{a}},\RX_{\rm{b}},\RX_{\rm{c}}$ and $\RX_{\rm{d}}$ be four uniformly random and
  independent bits. With $V=[5]$, define
  \begin{equation}
    \label{eq:receptacle}
    \begin{tikzpicture}[baseline=(current bounding box.center),ampersand replacement=\&]
      \matrix [matrix of math nodes, column sep=0em,inner sep=0.1em,outer sep=0] {
        \RZ_1:=\& ( \& \RX_{\rm{a}}, \& \RX_{\rm{b}}, \& \RX_{\rm{c}}, \& \RX_{\rm{d}} \& ) \\
        \RZ_2:=\& ( \& \RX_{\rm{a}}, \&  \& \RX_{\rm{c}} \& \& ) \\
        \RZ_3:=\& ( \& \RX_{\rm{a}}, \& \RX_{\rm{b}} \&  \&  \& ) \\
        \RZ_4:=\& ( \&  \& \RX_{\rm{b}}, \& \RX_{\rm{c}} \&  \& ) \\
        \RZ_5:=\& ( \&  \&  \&  \& \RX_{\rm{d}} \& ). \\
      };
    \end{tikzpicture}
  \end{equation}
  This is a hypergraphical source, illustrated in
  \figref{fig:receptacle} with edge set $E=\Set{\rm{a},\rm{b},\rm{c},\rm{d}}$, and edge
  function
  \begin{align*}
    `x(\rm{a})=\Set{1,2,3}, \quad`x(\rm{b})=\Set{1,3,4}, \\
    `x(\rm{c})=\Set{1,2,4}, \text{ and }`x(\rm{d})=\Set{1,5}.  
  \end{align*}
  The weight function $ c$ has $c(B)=1$ for $B$ equal to any of the subsets
  above, which form the support of $c$, i.e.,
  $\op{supp}(c)=\Set{\Set{1,2,3},\Set{1,3,4},\Set{1,2,4},\Set{1,5}}$.
\end{Example}

A simpler source model we shall also consider is the special hypergraphical source model when the hypergraph
  corresponds to a graph:
\begin{Definition}[\cite{nitinawarat10,nitinawarat-ye10}]
  \label{def:PIN}
  $\RZ_V$ is a \emph{pairwise independent
    network (PIN)} iff it is hypergraphical with edge function
  satisfying $\abs{`x(e)}=2$ for all $e\in E$.
\end{Definition}


\begin{Example}
\begin{figure}
    \centering 
    \begin{tikzpicture}
      \node[dot,label={[label distance=0em]above:{}}] (s) at(-2,0) {$1$}; 
      \node[dot,label={[label distance=0em]above:{}}] (a) at(0,0) {$2$} ; 
      \node[dot,label={[label distance=0em]above:{}}] (t) at(2,0) {$3$}; 
      \draw[edge] (-1.77,0) -- (-.25,0); 
      \node [] at (-1,0.20) {\scriptsize ${\rm{a}}$};     
      \draw[edge] (0.25,0.1) -- (1.77,0.1); 
      \node [] at (1,0.30) {\scriptsize ${\rm{b}}$};     
      \draw[edge] (0.25,-0.1) -- (1.77,-0.1); 
      \node [] at (1,-0.30) {\scriptsize ${\rm{c}}$};     
    \end{tikzpicture}
    \caption{The graphical representation of the PIN in~\eqref{eq:mot:src}.}
    \label{fig:mot}
  \end{figure}
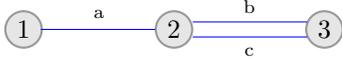
  With $V=[3]$, define
  \begin{equation}
    \label{eq:mot:src}
    \begin{aligned}
      \RZ_1 & :=\RX_{\rm{a}}\\
      \RZ_2 & :=(\RX_{\rm{a}},\RX_{\rm{b}},\RX_{\rm{c}})\\
      \RZ_3 & :=(\phantom{\RX_{\rm{a}},}\:\RX_{\rm{b}},\RX_{\rm{c}}). 
    \end{aligned}
  \end{equation}
  where $\RX_{\rm{a}},\RX_{\rm{b}},\RX_{\rm{c}}$ are independent
uniformly random bits. 
  The private source $(\RZ_1,\RZ_2,\RZ_3)$ is a PIN illustrated in 
  \figref{fig:mot} with edge set $E=\Set{\rm{a},\rm{b},\rm{c}}$, edge function $`x$ satisfying 
  \begin{align*}
    `x(\rm{a})=\Set{1,2}, `x(\rm{b})=`x(\rm{c})=\Set{2,3}, 
  \end{align*}
  and weight function 
  \begin{align*}
    c(\Set{1,2})=H(\RX_{\rm{a}})=1, c(\Set{2,3})=H(\RX_{\rm{b}})+H(\RX_{\rm{c}})=2, 
  \end{align*}
  and $0$ otherwise. Hence, the support of $c$ is $\op{supp}(c)=\Set{\Set{1,2},\Set{2,3}}$.
\end{Example}

\section{Preliminaries}
\label{sec:prelim}

If there is no limit on the public discussion rate, the secrecy
capacity, referred to as the \emph{unconstrained secrecy capacity}, is defined and characterized in
\cite{csiszar04} as
\begin{align}
  \CS(`8)&:=\lim_{R\to\infty}\CS(R) \label{eq:CS}\\
     &= \CS(\RCO) = H(\RZ_V)-\RCO\label{omni}
\end{align}
where $\RCO$ is the \emph{smallest rate of communication for
  omniscience}, characterized in \cite{csiszar04} by the linear
program
\begin{subequations}
  \label{eq:RCO}
  \begin{align}
    \RCO=\min_{r_V\in `R^V} r(V) \quad \text{such that}\\
    r(B)\geq H(\RZ_B|\RZ_{V`/B})\quad \forall B\subsetneq V,\label{eq:SW}
  \end{align}
\end{subequations}
where we have used for notational convenience that
\begin{align*}
  r(B):=\sum_{i\in B}r_i.
\end{align*}
The inequalities in~\eqref{eq:SW} consist of the usual Slepian-Wolf constraints for
source networks. The capacity-achieving scheme in \cite{csiszar04} requires all users
to recover the entire source $\RZ_V$ (i.e., attain \emph{omniscience}) by public discussion at the
smallest total rate $\RCO$, and then extract the secret key from
their recovered source at rate $H(\RZ_V)-\RCO$. Despite having exponentially many constraints, the linear program~\eqref{eq:RCO} can
be computed in (strongly) polynomial-time~\cite{chan11isit,milosavljevic11}, and hence, so can $ \CS(`8)$.

However, it was also mentioned in \cite{csiszar04} that the unconstrained
capacity can be attained by a possibly smaller discussion rate,
referred later in \cite{MKS16} as the \emph{communication complexity}
\begin{align}
	\RS 
	& = \min\{R\geq 0\mid \CS(R)=\CS(`8)\}\leq\RCO. \label{eq:RS}
\end{align}
For the PIN model, in particular, there is a protocol in \cite[Proof
of Theorem~3.3]{nitinawarat-ye10} that achieves the unconstrained
secrecy capacity~\cite[(15),(17)]{nitinawarat-ye10} possibly with
smaller discussion rate.
\begin{Proposition}[\mbox{\cite{nitinawarat-ye10,nitinawarat10}}]\label{pro:tree-packing}
  For a PIN,
  \begin{subequations}
    \begin{align}
      \CS(R)\geq \sum_{j\in [k]} `h_j \quad \forall R\geq (\abs{V}-2)\sum_{j\in [k]} `h_j
      \label{eq:tree-packing:rate}
    \end{align}
    where $k$ is a non-negative integer; $`h_j\in `R_+$ is a
    non-negative real number; $T_j:=(V,\mcE_j)$ is a spanning tree
    with edge set $\mcE_j\subseteq \Set{B\in \op{supp}(c)\mid |B|=2}$ satisfying
    \begin{align}
      \sum_{j\in [k]:B\in \mcE_j} `h_j \leq c(B)
      \kern1em \forall B\in {{V}\choose{2}}
      \label{eq:tree-packing:cons}
    \end{align}
  \end{subequations}
  Furthermore, the lower bound in \eqref{eq:tree-packing:rate}
  achieves the unconstrained secrecy capacity.
\end{Proposition}
Note that the feasible solutions $\Set{(`h_j,T_j)\mid j\in [k]}$ to the lower bound in
\eqref{eq:tree-packing:rate} are called \emph{fractional tree
  packings} because the constraint~\eqref{eq:tree-packing:cons} requires the
total weights $\sum_{j\in [k]:B\in \mcE_j} `h_j $ of all the trees
covering each set $B$ to not exceed the
weight $c(B)$. The achieving scheme is therefore
called the \emph{tree-packing protocol}. The
unconstrained secrecy capacity is the \emph{fractional tree packing number}.

It was left as an open problem in \cite{nitinawarat10}
whether the tree-packing protocol achieves the communication complexity
$\RS$. 
One may further ask whether the scheme achieves the constrained
secrecy capacity $\CS(R)$ for all $R\geq 0$.
We resolve this in the
affirmative in Theorem~\ref{thm:PIN} by providing a matching converse. This idea can be
motivated more concretely with the following example.
  
\begin{Example}
  \label{eg:mot}
  Consider the PIN model defined in \eqref{eq:mot:src}.
  If user 2 reveals $\RF:=\RX_{\rm{a}}\oplus\RX_{\rm{b}}$ in public so
  that everyone can observe it, then user 3 can recover $\RX_{\rm{a}}$
  as $\RF\oplus\RX_{\rm{b}}$. $\RK:=\RX_{\rm{a}}$ is a secret key bit
  generated by the public discussion $\RF$ because not only is $\RK$ 
  recoverable by all users, with the recoverability constraint~\eqref{eq:recover} being satisfied, but it is also uniformly random and
  independent of the public discussion $\RF$, thus satisfying the secrecy
  constraint~\eqref{eq:secrecy}.

  The above secret key agreement scheme is indeed a tree
  packing protocol. There is only one possible spanning tree, namely $T_1=(V,\mcE_1)$
  with $\mcE_1=\Set{\Set{1,2},\Set{2,3}}$. To satisfy the weight constraint~\eqref{eq:tree-packing:cons}, we can set $\eta_1\leq 1 =c(\Set{1,2})\leq c(\Set{2,3})$.
  Hence, it follows from \eqref{eq:tree-packing:rate} that, for $\eta_1\in [0,1]$,
  \begin{align*}
    \CS(R)&\geq \eta_1 \quad \forall R\geq \eta_1,
  \end{align*}
  or equivalently
   \begin{align}
   \CS(R) &\geq \min\Set{R,1}.\label{eq:CSR:mot}
  \end{align}
  It is easy to see that the capacity cannot exceed $1$~bit since user 1 observes at most $1$~bit in private, and $1$~bit
  of secret key is achievable by the above discussion scheme. The
  smallest rate of communication for omniscience is $\RCO=2$ because
  there are $H(\RZ_V)=3$~bits of randomness in the source but user~$1$
  only gets to observe $1$~bit in private. It can be checked that the
  formula~\eqref{omni} relating $\CS(`8)$ and $\RCO$ holds, and that the
  linear program~\eqref{eq:RCO} for $\RCO$ is solved by the rate tuple
  $(r_1,r_2,r_3)=(0,1+\epsilon,1-\epsilon)$ for any $\epsilon \in [0,1]$. The bound~\eqref{eq:RS} on communication complexity
  is $\RS\leq \RCO=2$. However, this bound is loose because the
  earlier capacity-achieving discussion $\RF$ is only $1$~bit, i.e.,
  we have $\RS\leq 1< \RCO$. It can be shown that the best existing lower bound
  from~\cite{MKS16,chan16itw} is $\RS\geq 0$, which is trivial. Hence,
  the existing result are not sufficient to characterize $\RS$, let alone
  the constrained secrecy capacity $\CS(R)$.

  It turns out that the lower bound \eqref{eq:CSR:mot} on $\CS(R)$ for
  the current example is tight, which implies $\RS=1$. Proving the reverse
  inequality is non-trivial
  and is the motivation of the techniques
  introduced in this work.
\end{Example}

\section{Main results}
\label{sec:main}

Unless otherwise stated, all the results apply to the hypergraphical
source model in Definition~\ref{def:hyp} with the additional
assumptions stated after that. We will also consider the non-trivial
case involving $\abs{V}>2$ users. For ease of understanding, we will
present the most general result towards the end of this section, after introducing some of its simpler
variants which already give tight characterizations of the capacities for simple hypergraphical sources.

\subsection{Edge-partition bound}
\label{sec:EP}

Let $\Pi'(V)$ be the collection of partitions of $V$ into at least two
non-empty disjoint sets.

\begin{Theorem}[EP bound] For any partition $\mcP\in \Pi'(V)$, an
  upper bound on the
  constrained secrecy capacity~\eqref{eq:CSR} is given by
  \label{thm:EP}
  \begin{subequations}
    \label{eq:EP}
    \begin{align}
      & [1-`a(\mcP)] \CS(R) \leq `a(\mcP) R, \quad \text{where}\\
      &`a(\mcP) := \frac{\max_{e\in E} \abs{\Set{C\in \mcP\mid `x(e)\cap
        C\neq `0}}-1}{\abs{\mcP}-1}.\label{eq:`a}
    \end{align}
  \end{subequations}
\end{Theorem}
This is called the \emph{edge-partition (EP) bound}.
\begin{Proof}
  See Appendix~\ref{sec:proof:EP}
\end{Proof}

Note that we did not incorporate the obvious upper bound $\CS(R)\leq \CS(`8)$ into
\eqref{eq:EP} to avoid distraction. This obvious upper bound will also
be implicit in the subsequent results. The name ``edge-partition bound''  is because the critical component 
$`a(\mcP)$ of the bound is obtained by partitioning the edges of the
hypergraph. More precisely, in the numerator of $`a(\mcP)$ in \eqref{eq:`a}, the expression $\Set{C\in \mcP\mid `x(e)\cap
  C\neq `0}$ is the collection of subsets in the partition $\mcP$
that intersects the incident nodes $`x(e)$ of an edge $e$. The size of this collection minus $1$
is the number of times $\mcP$ cuts across the edge $e$. Therefore, the
numerator of $`a(\mcP)$ is the maximum number of times $\mcP$ can
cut across an edge of the hypergraphical source. The denominator
is the number of cuts across the entire vertex set $V$. Hence, 
$`a(\mcP)$ is a ratio no larger than $1$, with equality if there is an edge $e$ that
covers the entire vertex set, i.e., $`x(e)=V$.

An example that illustrates the EP bound is as follows.

\begin{Example}
  \label{eg:receptacle}
  Consider the hypergraphical source defined in
  \eqref{eq:receptacle}. (See \figref{fig:receptacle}.)
  We will compute the tightest EP bound among all possible values of
  the partition $\mcP\in \Pi'(V)$. Consider the case $\abs{\mcP}=5$,
  namely, the singleton partition
  $\mcP=\Set{\Set{1},\Set{2},\Set{3},\Set{4},\Set{5}}$. The following
  matrix lists the non-zero values of the indicator function $\chi\Set{`x(e)\cap C\neq `0}$ for different edges $e\in E$ and blocks $C\in \mcP$ of the partition.
  \begin{center}
    \begin{tikzpicture}
      \matrix [matrix of math nodes] {
        |(E)| & \rm{a} & \rm{b} & \rm{c} & |(d)| \rm{d} \\
        |(C1)| \Set{1} & 1 & 1 & 1 & 1 \\
        \Set{2} & 1 &  & 1 &  \\
        \Set{3} & 1 & 1 & & \\
        \Set{4} &  & 1 & 1 & \\
        |(5)| \Set{5} &  &  & & |(5d)| 1\\
        |(sum)|\text{sum} & 3 & 3 & 3 & |(last)| 2\\
      };
      \path (5) to node (5sum) {} (sum);
      \draw (5sum-|sum.west) -- (d.east|-5sum);
    \end{tikzpicture}
  \end{center}
  The last row gives the column sums, which corresponds to the values of $\abs{\Set{C\in \mcP\mid `x(e)\cap C\neq `0}}$. The maximum value is $3$, and so
  \begin{align*}
    `a(\mcP) = \frac{3-1}{5-1} = \frac12\quad \forall \mcP\in \Pi'(V):\abs {\mcP}=5.
  \end{align*}

  For $\mcP\in \Pi'(V)$ with $\abs{\mcP}=4$, there are ${4 \choose 2}=6$ possible partitions. The values of $`a(\mcP)$ can be computed similarly. It can be checked that $\max_{e\in E} \abs{\Set{C\in \mcP\mid `x(e)\cap C\neq `0}} \geq 3$ and so
  \begin{align*}
    `a(\mcP) \geq \frac{3-1}{4-1} = \frac23\quad \forall \mcP\in \Pi'(V):\abs {\mcP}=4.
  \end{align*}
  This gives a looser EP bound compared to the previous case with
  $\abs{\mcP}=5$ as it can be observed from the EP bound~\eqref{eq:EP}
  that a smaller value of $`a(\mcP)$ gives a tighter bound.

  Similarly, for $\mcP\in \Pi'(V)$ with $\abs{\mcP}\leq 3$, it can be shown that
  $\abs{\Set{C\in \mcP\mid `x(e)\cap C\neq `0}}\geq 2$ and so
  $`a(\mcP)\geq \frac12$, which again cannot give a better EP bound
  than the case with $\abs{\mcP}=5$. Hence, with $`a(\mcP)=\frac12$,
  we have 
  the tightest EP bound
  \begin{align*}
    `1(1-\frac12`2)\CS(R)&\leq \frac12 R
  \end{align*}
  which gives $\CS(R)\leq R$.
\end{Example}

Although the EP bound can be computed efficiently given a
particular choice of the partition $\mcP$, it
is unclear how to efficiently compute the optimal partition $\mcP$ that gives
the tightest EP bound. As we
shall see in next section, the EP bound is also loose for the above example. 

Nevertheless, when restricted to the PIN model in
Definition~\ref{def:PIN}, the optimal partition turns out to be the
partition $\Set{\Set{i}\mid i\in V}$ into singletons. The tightest EP bound gives a complete and surprisingly simple characterization of
the constrained secrecy capacity, which is also achieved by the
tree-packing protocol described in Proposition~\ref{pro:tree-packing}
and illustrated in Example~\ref{eg:mot}.

\begin{Theorem}
  \label{thm:PIN}
  For the PIN model, the constrained
  secrecy capacity is
  \begin{align}
    \CS(R) = \min \Set*{\frac{R}{\abs{V}-2},\CS(`8)}\label{eq:CSR:PIN}
  \end{align}
  for the case of interest when $\abs{V}\geq 3$. It follows that the
  communication complexity defined in \eqref{eq:RS} is $\RS=(\abs{V}-2)\CS(`8)$.
\end{Theorem}
\begin{Remark}
The optimal tradeoff is irrelevant to the topology of the PIN model, and is characterized simply by the size of the network.
\end{Remark}
\begin{Proof}
  Note that the lower bound $\geq$ of~\eqref{eq:CSR:PIN} directly follows from
  \eqref{eq:tree-packing:rate} in
  Proposition~\ref{pro:tree-packing}. Furthermore, $\CS(`8)$
  equals the fractional tree-packing number.

  The converse follows from~\eqref{eq:EP} with $\mcP=\Set{\Set{i}\mid
    i\in V}$. More precisely, the maximization in the numerator of
  $`a(\mcP)$ is always equal to $2$ as it is the number of incident
  nodes of an edge. Hence, $\alpha(\mcP)=\frac{1}{\abs{V}-1}$, and so,
  the EP bound gives
  \begin{align*}
    (\abs{V}-2) \CS(R)\leq R,
  \end{align*}
  which completes the proof of
  \eqref{eq:CSR:PIN}.
  The formula for $\RS$ is obtained easily by equating the two terms
  in the minimization in \eqref{eq:CSR:PIN}.
\end{Proof}

For the PIN defined in \eqref{eq:mot:src}, for instance, the bound~\eqref{eq:CSR:mot} of $\CS(R)$ is the precise characterization
given by the above equation~\eqref{eq:CSR:PIN}. As discussed, the tradeoff is irrelevant to the topology. To illustrate, another $3$-user
example is as follows.

\begin{Example}
\label{eg:triangle}
  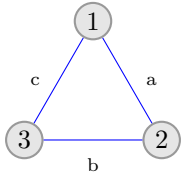
\begin{figure}
    \centering
    \begin{tikzpicture}[x=.6em,y=.6em,>=latex]
      \foreach \x/\angle/\lb in {3/210/{},1/90/{},2/-30/{}}
      {
	\path (\angle:5) node (\x) [dot,label={[label distance=0em]\angle:{\lb}}] {$\x$};
      }
      \foreach \x/\y/\lp/\lb in {1/2/right/${\rm{a}}$,
	2/3/below/${\rm{b}}$,
	1/3/left/${\rm{c}}$}
      \draw[edge] (\x) to node [label=\lp:{\scriptsize\lb}]{} (\y);
    \end{tikzpicture}
    \caption{The triangle PIN defined in~\eqref{eq:triangle}.}
    \label{fig:triangle}
  \end{figure}

  Consider a triangle PIN with
  $V=[3]$ and
  \begin{equation}
    \label{eq:triangle}
    \begin{aligned}
      \RZ_1&:=(\RX_{\rm{a}},\kern1.6em\RX_{\rm{c}})\\
      \RZ_2&:=(\RX_{\rm{a}},\RX_{\rm{b}}\kern1.7em)\\
      \RZ_3&:=(\kern1.7em\RX_{\rm{b}},\RX_{\rm{c}})
    \end{aligned}
  \end{equation}
  where $\RX_{\rm{a}},\RX_{\rm{b}},\RX_{\rm{c}}$ are independent
uniformly random bits. This is a PIN with correlation represented by a
triangle as shown in \figref{fig:triangle}. It follows from
\eqref{eq:CS}, \eqref{eq:RCO} and \eqref{eq:RS} that 
$$\CS(`8)=\RCO=1.5\geq \RS.$$
The capacity $1.5$ is achievable by the 
tree-packing protocol with the following fractional solution
\begin{align*}
  \mcE_1&=\Set{\Set{1,2},\Set{2,3}} & \eta_1 &=\tfrac12 \\
  \mcE_2&=\Set{\Set{1,2},\Set{1,3}} & \eta_2 &=\tfrac12 \\
  \mcE_3&=\Set{\Set{2,3},\Set{1,3}} & \eta_3 &=\tfrac12.
\end{align*}
Applying \eqref{eq:CSR:PIN}, we have
\begin{align*}
  \CS(R)=\min\Set{R,1.5}\quad \text{ and }\quad
  \RS=1.5=\RCO.
\end{align*}
The tradeoff is the same as the PIN defined in~\eqref{eq:mot:src}. Note that unlike Example~\ref{eg:mot}, the characterization of
$\RS$ for this example can be obtained using an existing technique in \cite{chan16itw}
by showing the \emph{optimality of omniscience}.
\end{Example}

\subsection{Vertex-packing bound}
\label{sec:VP}

Although the EP bound is tight for the PIN model, with
$\mcP=\Set{\Set{i}\mid i\in V}$, it can be loose in general. In
particular, the following result will give a tighter upper bound on $\CS(R)$ 
for the hypergraphical source considered earlier in Example~\ref{eg:receptacle}.

\begin{Theorem}[VP bound]
  \label{thm:VP}
For any $R\geq 0$,  
  \begin{subequations}
    \label{eq:VP}
    \begin{align}
      &(`t-1)\CS(R)\leq R,\quad\text{where}\\
      &`t:=\max_{u_V\in `R_+^V:u(`x(e))\leq 1,\forall e\in E} u(V).\label{eq:`t}
    \end{align}
  \end{subequations}
\end{Theorem}
This is called the \emph{vertex-packing (VP) bound}.

\begin{Proof}
  See Appendix~\ref{sec:proof:VP}.
\end{Proof}

Note that the name of the bound comes from the fact that the feasible solution $u_V$ to the above linear
program~\eqref{eq:`t} is a fractional collection of the vertices that
can be packed into the hyperedges. We will show below that the VP
bound can be tighter than the tightest EP bound $\CS(R)\leq R$
obtained in Example~\ref{eg:receptacle}. Note also that the VP bound
can be computed more efficiently than the tightest EP bound. 

\begin{Example}
  \label{eg:VP}
  Consider the previous hypergraphical source defined in \eqref{eq:receptacle}. The constraints of \eqref{eq:`t} are
  \begin{align*}
    \begin{tikzpicture}[baseline=(current bounding box.center),ampersand replacement=\&]
      \matrix [matrix of math nodes, column sep=0em,inner sep=0.1em,outer sep=0] {
        u_1\& +u_2 \& +u_3 \& \& \& \leq 1\\
        u_1\&  \& +u_3 \& +u_4 \& \& \leq 1\\
        u_1\& +u_2 \&  \& +u_4\& \& \leq 1\\
        u_1\& \& \& \& +u_5 \& \leq 1\\
      };
    \end{tikzpicture}
  \end{align*}
  The optimal solution to \eqref{eq:`t} is given uniquely by
  \begin{align*}
    \bM u_1 \\ u_2 \\ u_3 \\ u_4 \\ u_5 \eM
    &= \bM 0\\ \frac12 \\ \frac12 \\ \frac 12 \\ 1\eM,
    \text{ which achieves }`t=\sum_{i=1}^5 u_i=\frac52.
  \end{align*}
  Hence, by \eqref{eq:VP}, $`1(\frac52-1`2)\CS(R)\leq R$, or
  equivalently, $\CS(R)\leq \frac23 R$. This bound is not only tighter
  than the tightest EP bound $\CS(R)\leq R$, but it can also be shown to be
  achievable, i.e., it can be shown that $\CS(R)=\frac23R$ up to the unconstrained capacity $\CS(`8)=1$.
 Consider two independent realizations of the source, i.e., let $\RX_{\rm{a}t},\RX_{\rm{b}t},\RX_{\rm{c}t},\RX_{\rm{d}t}$ be the independent bits at time $t\in\Set{1,2}$. User~$1$ reveals $\RF_1=(\RX_{\rm{a}1}\oplus\RX_{\rm{d}1}, \RX_{\rm{c}1}\oplus\RX_{\rm{d}2}, \RX_{\rm{a}1}\oplus\RX_{\rm{b}1}\oplus\RX_{\rm{d}2})$ in public. Then, the users can agree on $2$~bits of secret key, namely $\RK=(\RX_{\rm{d}1},\RX_{\rm{d}2})$, which are independent of the discussion.  Since $2$~bits of secret key can be agreed by $3$~bits of discussion for every $2$~units of time, we have $\CS(\frac{3}{2})\geq \frac{2}{2}=1$. By the usual time-sharing argument, $\CS(R)\geq \min\Set{\frac23R,1}$ as desired. Therefore, the VP bound is tight for this example.
\end{Example}

Note that the
VP bound is not always better than the EP bound, i.e., it is possible for the EP bound to be strictly tighter
than the VP bound, as the following example shows. 

\begin{Example}
Consider the PIN model on a complete graph, i.e., with $E={{V}\choose{2}}$. (See the triangle PIN in~Example~\ref{eg:triangle} for $\abs{V}=3$.) Note that
\begin{align*}
`t=\max_{u_V\in `R_+^V:u_i+u_j \leq 1, \forall i,j \in V} u(V)=\frac{\abs{V}}{2}.
\end{align*} 
Then, by~\eqref{eq:VP}, the VP bound is
\begin{align*}
`1[\frac{\abs{V}}{2}-1`2]\CS(R)\leq R
\end{align*}
which is worse than the EP bound $`1[\abs{V}-2`2]\CS(R)= R$ in the non-trivial case $\abs{V}>2$.      
\end{Example}

\subsection{Lamination Bound}
\label{sec:lb}

It is possible that both the EP and VP bounds are loose. 
Indeed, the bounds can be unified and improved to a more
general bound, called lamination bound below:

\begin{Theorem}[Lamination Bound]
\label{thm:lb}
	For all $R\geq 0$,
	\begin{subequations}
	\label{eqs:lb}
	\begin{align}
		`b(\pi) \CS (R) \leq `1[`g(`l,`p,`r) - `b(\pi)`2] R \label{eq:lb}
	\end{align}	
	where 
	\begin{align}
		`b(\pi) &:= \min_{e\in E} \sum_{B\in 2^V`/\Set {`0,V}: `x(e)\subseteq B} \pi(B)\label{eq:`b}\\
		`g(`l,`p,`r) &:= `r+\sum_{B\in 2^V`/\Set {`0,V}} `1[`p(B)-`r`l(B)`2]\label{eq:`g}.
	\end{align}
	\end{subequations}
        The parameter $`r\geq 0$ and set functions $`l,`p:2^V\to
        `R_+$ are chosen such that
        \begin{subequations}
          \label{eq:lb:constraints}
	\begin{align}
		&\sum_{B\subseteq 2^V`/\Set {`0,V}:i\in B} `l(B) \leq 1 \kern1em \forall i\in V \label{eq:FP}\\
		&`p(B)-`r`l(B) \geq 0 \kern1em \forall B\in 2^V`/\Set{`0,V}\label{eq:rou}.
	\end{align}
	and so $`l$ is a fractional packing of $V$ according to
        \eqref{eq:FP}.
        \end{subequations}
\end{Theorem}

\begin{Proof}
  See Appendix~\ref{sec:proof:lb}.
\end{Proof}

\begin{Corollary}
\label{cor:EP:VP}
The lamination bound~\eqref{eqs:lb} covers the EP bound~\eqref{eq:EP} and VP bound~\eqref{eq:VP} as special cases.
\end{Corollary}
\begin{Proof}
  See Appendix~\ref{sec:proof:EP:VP}.
\end{Proof}

The following example illustrates the above lamination bound and shows
that it can be strictly better than the EP and VP bounds.
\begin{Example}
  \label{eg:lb}
  Consider the following hypergraphical source, illustrated in \figref{fig:scoop}:
  \begin{equation}
    \label{eq:scoop}
    \begin{tikzpicture}[baseline=(current bounding box.center),ampersand replacement=\&]
      \matrix [matrix of math nodes, column sep=0em,inner sep=0.1em,outer sep=0] {
        \RZ_1:=\& ( \& \RX_{\rm{a}}, \& \RX_{\rm{b}}, \& \RX_{\rm{c}} \&  \& ) \\
        \RZ_2:=\& ( \& \RX_{\rm{a}}, \&  \& \RX_{\rm{c}}, \& \RX_{\rm{d}} \& ) \\
        \RZ_3:=\& ( \& \RX_{\rm{a}}, \& \RX_{\rm{b}} \&  \&  \& ) \\
        \RZ_4:=\& ( \&  \& \RX_{\rm{b}}, \& \RX_{\rm{c}} \&  \& ) \\
        \RZ_5:=\& ( \&  \&  \&  \& \RX_{\rm{d}} \& ) \\
      };
    \end{tikzpicture}
  \end{equation}
  Compared to the source defined in \eqref{eq:receptacle} and illustrated in \figref{fig:receptacle}, the difference is that the edge $\rm{d}$ connects node~$5$ to node~$2$ instead of node~$1$.
  
  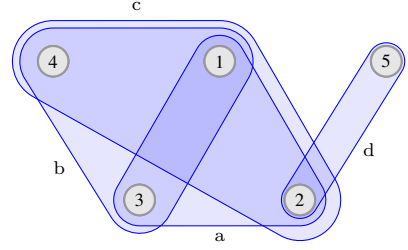
\begin{figure}
    \centering
    \def\u{1em}
    \begin{tikzpicture}[x=.6em,y=.6em,>=latex]
      \scriptsize
      \path (90:5*\u) node (1) [dot] {1};
      \path (-30:5*\u) node (2) [dot] {2};
      \path (210:5*\u) node (3) [dot] {3};
      \path (1) +(-9*\u,0*\u) node [dot] (4) {4};
      \path (1) +(9*\u,0*\u) node [dot] (5) {5};

      \draw[hyperedge] \convexpath{1,2,3}{1.4*\u};
      \node at ($0.5*(hullcoord2)+0.5*(hullcoord3)$) [yshift=-2*\u] {$\rm{a}$};
      \draw[hyperedge] \convexpath{1,3,4}{1.8*\u};
      \node at ($0.5*(hullcoord2)+0.5*(hullcoord3)$) [xshift=-2*\u,yshift=-2*\u] {$\rm{b}$};
      \draw[hyperedge] \convexpath{1,2,4}{2.2*\u};
      \node at ($0.5*(hullcoord0)+0.5*(hullcoord1)$) [xshift=0*\u,yshift=3*\u] {$\rm{c}$};
      \draw[hyperedge] \convexpath{2,5}{1*\u};
      \node at ($0.5*(hullcoord1)+0.5*(hullcoord2)$) [xshift=1.4*\u,yshift=-1*\u] {$\rm{d}$};

      \foreach \x in {1,2,...,5}
      \node at (\x) [dot] {\x};

    \end{tikzpicture}
    \caption{The hypergraphical source defined in \eqref{eq:scoop}.}
    \label{fig:scoop}
  \end{figure}
  
We can calculate the tightest EP bound as in Example~\ref{eg:receptacle}: 
\begin{itemize}
\item For $\abs{\mcP}=5, \alpha(\mcP)=\frac{3-1}{5-1}=\frac{1}{2}$;
\item For $\abs{\mcP}=4, \alpha(\mcP)\geq \frac{3-1}{4-1}=\frac{2}{3}$;
\item For $\abs{\mcP}=3, \alpha(\mcP)\geq\frac{2-1}{3-1}=\frac{1}{2}$;
\item For $\abs{\mcP}=2, \alpha(\mcP)\geq\frac{2-1}{2-1}=1$.
\end{itemize}
Therefore, the tightest EP bound~\eqref{eq:EP} is $\CS(R)\leq R$, which was given by the smallest $\alpha(\mcP)=\frac12$.

For the VP bound, the constraints of~\eqref{eq:`t} are
  \begin{align*}
    \begin{tikzpicture}[baseline=(current bounding box.center),ampersand replacement=\&]
      \matrix [matrix of math nodes, column sep=0em,inner sep=0.1em,outer sep=0] {
        u_1\& +u_2 \& +u_3 \& \& \& \leq\& 1\\
        u_1\&  \& +u_3 \& +u_4 \& \& \leq\& 1\\
        u_1\& +u_2 \&  \& +u_4\& \& \leq\& 1\\
        \& \kern0.8em u_2\& \& \& +u_5 \& \leq\& 1\\
      };
    \end{tikzpicture}
  \end{align*}
It can be show that $`t=\sum_{i=1}^5 u_i=2$, with $u_1=u_5=1$ and $u_2=u_3=u_4=0$. Therefore, the VP bound~\eqref{eq:VP} is $\CS(R)\leq R$.

Computing the tightest lamination bound is not easy due to its
generality. For this example, however, the lamination bound turns out
to be achievable (and therefore tight) for following choice of parameters:
\begin{align*}
  `r&=\frac{20} {3}\\
  `l(B)&=
         \begin{cases} 
           0.6, &B\in\Set{\Set{1,3,4},\Set{2,5}}\\
           0.15, &B\in\Set{\Set{1,2,3,5},\Set{1,2,4,5}}\\
           0.1, &B=\Set{1,2,3,4}\\
           0, & \text{otherwise}.
         \end{cases}\\
  `p(B)&=
         \begin{cases} 
           4, &B\in\Set{\Set{1,3,4},\Set{2,5},\Set{1,2,3,5},\Set{1,2,4,5}}\\
           8, &B=\Set{1,2,3,4}\\
           0, & \text{otherwise}.
         \end{cases}
\end{align*}  
It is straight-forward to check that $`r,`l$ and $ `p$ satisfy the
constraints~\eqref{eq:lb:constraints}. By~\eqref{eq:`b}
and~\eqref{eq:`g}, we have $`b(`p)=12, `g(`l,`p,`r)=20$. The
lamination bound~\eqref{eq:lb} is $\CS(R)\leq \frac23 R$, which is
strictly better than the EP and VP bounds.

To show that the bound is
tight up to the unconstrained secrecy capacity, consider $n=2$ independent realizations of the private source,
i.e., let $\RX_{\rm{a}t},\RX_{\rm{b}t},\RX_{\rm{c}t},\RX_{\rm{d}t}$ be the independent
bits at time $t\in\Set{1,2}$.  If user 1 and user 2 reveal in public
$\RF_1=\RX_{\rm{a}1}\oplus\RX_{\rm{b}1}\oplus\RX_{\rm{c}1}$ and
$\RF_2=(\RX_{\rm{a}1}\oplus\RX_{\rm{d}2}, \RX_{\rm{c}1}\oplus\RX_{\rm{d}1})$,
respectively, then all users can recover $
\RX_{\rm{a}1},\RX_{\rm{b}1},\RX_{\rm{c}1},\RX_{\rm{d}1},\RX_{\rm{d}2}$ perfectly. Let
$\RK=(\RX_{\rm{d}1},\RX_{\rm{d}2})$ be the secret key, which is independent of
the discussion $(\RF_1,\RF_2)$. Since the users can agree on $2$~bits
of secret key by $3$~bits of discussion in $2$ time units, the constrained secrecy capacity is $\CS(R)\geq \min\Set{\frac23R,1}$.
\end{Example}

\section{Conclusion}
\label{sec:conclusion}

In this work, we give explicit upper bounds on the maximum secret key rate achievable by any given total public discussion rate. The focus is on the multiterminal setting, where we specialize to the hypergraphical source model to circumvent the difficulties of the $2$-user case. We exploited the independence of the edge variables to derive the bounds using the lamination procedure from Edmonds' greedy algorithm in submodular function optimization. The bounds include
\begin{itemize}
\item the EP bound~\eqref{eq:EP}, which is shown to be tight for the PIN model and therefore gives a surprising simple characterization of the capacity achievable by tree packing;
\item the VP bound~\eqref{eq:VP}, which can sometimes give a tighter bound than the EP bound as illustrated by Example~\ref{eg:VP}; and
\item the lamination bound~\eqref{eqs:lb}, which generalizes and strictly improves both the EP and VP bounds as shown by Example~\ref{eg:lb}.
\end{itemize}
These bounds do not involve any auxiliary random variables, making them easier to compute than the usual single-letter solutions. Furthermore, the strongest lamination bound appears to be tight for all the examples we have constructed. 

Nevertheless, we suspect that the lamination bound may be further improved. An
alternative proof of upper bound for Example~\ref{eg:lb} is given in
Appendix~\ref{sec:challenge}. The alternative proof appears to be more elegant,
as it applies the lamination procedure two
times in a row to give the desired bound $\CS(R)\leq \frac23R$ more
naturally. However, it is unclear how to give an analytical form to
the bound obtained from iterative lamination. We also do
not yet have an example to show that iterative lamination gives a
strictly tighter bound. 

As in \cite{chan17oo}, there are new challenges when extending the results to a more general setting involving trusted/untrusted helpers~\cite{csiszar04} and silent active users~\cite{amin10a}. Even for the simple PIN model with some trusted helpers, the tree-packing protocol extended using Steiner trees is not optimal because it may not attain the unconstrained capacity~\cite{nitinawarat10}. There is a more general network coding approach in~\cite{chan11isit,chan12ud} for secret key agreement, and we have used such an approach in constructing the optimal secret key agreement schemes for the hypergraphical sources in Example~\ref{eg:VP} and \ref{eg:lb}. We believe this approach is optimal for general hypergraphical sources, but a rigorous proof remains elusive even for the basic case without helpers.

Another interesting question is to efficiently generate independent groupwise
secret keys instead of a global secret key. Note that the groupwise
secret keys form a hypergraphical source, with each group
represented by an edge with weight equal to
the total secret key rate of that group. Essentially, the question
of interest is whether one can convert from one hypergraph (the original
source) to
another (the groupwise secret keys), and if so, how much effort
(discussion rate) is required. It can be shown that public
discussion is an irreversible process, in a way, like the second law of
thermodymics. It therefore defines a partial ordering of weighted hypergraphs, with the PIN models at one end and the trivial
hypergraphs with edges covering all vertices in the other end. This
combinatorial structure may be interesting and useful in its own right.

\appendices

\makeatletter
\@addtoreset{equation}{section}
\renewcommand{\theequation}{\thesection.\arabic{equation}}
\renewcommand{\theparentequation}{\thesection.\arabic{parentequation}}
\@addtoreset{Theorem}{section}
\renewcommand{\theTheorem}{\thesection.\arabic{Theorem}}
\@addtoreset{Lemma}{section}
\renewcommand{\theLemma}{\thesection.\arabic{Lemma}}
\@addtoreset{Corollary}{section}
\renewcommand{\theCorollary}{\thesection.\arabic{Corollary}}
\@addtoreset{Example}{section}
\renewcommand{\theExample}{\thesection.\arabic{Example}}
\@addtoreset{Remark}{section}
\renewcommand{\theRemark}{\thesection.\arabic{Remark}}
\@addtoreset{Proposition}{section}
\renewcommand{\theProposition}{\thesection.\arabic{Proposition}}
\@addtoreset{Definition}{section}
\renewcommand{\theDefinition}{\thesection.\arabic{Definition}}
\@addtoreset{Subclaim}{Theorem}
\renewcommand{\theSubclaim}{\theLemma.\arabic{Subclaim}}
\makeatother

\section{Proof of EP bound in Theorem~\ref{thm:EP}}
\label{sec:proof:EP}

\def\u{1em}
\def\MarkLt{0.4*\u}
\def\MarkSep{0.2*\u}
\tikzset{
	TwoMarks/.style={
		postaction={decorate,
			decoration={
				markings,
				mark=at position #1 with
				{
					\begin{scope}[xslant=0.2]
						\draw[line width=\MarkSep,white,-] (0pt,-\MarkLt) -- (0pt,\MarkLt) ;
						\draw[-] (-0.5*\MarkSep,-\MarkLt) -- (-0.5*\MarkSep,\MarkLt) ;
						\draw[-] (0.5*\MarkSep,-\MarkLt) -- (0.5*\MarkSep,\MarkLt) ;
					\end{scope}
				}
			}
		}
	},
	TwoMarks/.default={0.5},
}

\tikzstyle{point}=[minimum size=0em,inner sep=0, outer sep=0em]
\tikzstyle{htag}=[draw,rectangle,minimum width=.4*\u,minimum height=0,inner sep=0, outer sep=0em]
\tikzstyle{vtag}=[draw,rectangle,minimum height=.4*\u,minimum width=0,inner sep=0, outer sep=0em]
\tikzstyle{aline}=[dashed,gray]
\tikzstyle{bar}=[fill=cyan!10!white!90,draw=cyan]

\begin{figure*}
	\begin{center}
	\subcaptionbox{$`m^*$ in general~\eqref{eq:`m*}.\label{fig:greedy:general}}{
	\def\ux{3em}
	\def\uy{2.5em}
	\tikzstyle{point}=[minimum size=0em,inner sep=0, outer sep=0em]
	\tikzstyle{htag}=[draw,rectangle,minimum width=.4*\u,minimum height=0,inner sep=0, outer sep=0em]
	\tikzstyle{vtag}=[draw,rectangle,minimum height=.4*\u,minimum width=0,inner sep=0, outer sep=0em]
	\tikzstyle{aline}=[dashed,gray]
	\tikzstyle{bar}=[fill=cyan!10!white!90,draw=cyan]
	\begin{tikzpicture}[>=latex]
	\scriptsize
	\node [point] (0) at (0,0) {};
	
	\foreach \j/\p/\w/\wl/\sl in {%
		1/0/6/{$w_{s_1}$}/{$s_1$},%
		3/1.2/4/{$w_{s_j}$}/{$s_{j}$},%
		4/2.2/3/{$w_{s_{j+1}}$}/{$s_{j+1}$},%
		7/3.4/1/{$w_{s_k}$}/{$s_k$}%
	}
	{
		\draw[bar] (\p*\ux,0) -- +(0,\w*\uy) node [point] (\j) {} -- +(1*\ux,\w*\uy)  -- +(1*\ux,0) -- cycle;
		\path (\j)  -- (0|-\j) node (w\j) [htag,label={[label distance=0.1*\u]180:\wl}] {};
		\draw[aline] (\j) -- (w\j.center);
		\path (\j|-0) to node [vtag,label={[label distance=0.4*\u]-90:\sl}] {} +(1*\ux,0);
	};
	
	\foreach \j/\lb in {%
		3/$\kern3.2em S_j:=\Set{s_{j'}\mid  j'\leq j}$%
		,7/$S_k=S$%
	}
	\draw[decorate, decoration={brace, amplitude=3pt,mirror,raise=0.1*\u}] (\j) +(\ux,0)  node (S\j) {}  -- (0|-S\j) node [midway,above,yshift=0.2*\u] {\contour{white}{\lb}}; 
	

	\foreach \j/\lb in {%
		3/$`m^*(S_j):=w_{s_j}-w_{s_{j+1}}$%
	}
	\draw[|<->|] (\j) +(1.1*\ux,0) -- +(1.1*\ux,-\uy) node [midway,right,xshift=0.2*\u] {\contour{white}{\lb}} ; 
	
	\draw[|<->|] (7) +(1.1*\ux,0) -- +(1.1*\ux,-\uy) node [midway,right,xshift=0.2*\u] {$`m^*(S):=w_{s_k}\kern-1em $} ; 
	
	\path (0) to node  {}  (4.8*\ux,0) node (s) [point,label={[label distance=0.1*\u]right:$s\in S$}] {};
	\path (0) -- (0,7*\uy) node (w) [point,label={[label distance=0.1*\u]above:$w_s$}] {};
	
	\draw[-,TwoMarks=0.675] (0)-- (4-|0);
	\draw[->,TwoMarks] (4-|0)-- (w);
	\path (0) -- +(\ux,0) node [point] (bx1) {};
	\draw[-] (0) -- (bx1);
	\draw[-,TwoMarks] (bx1)-- (3|-0);
	\path (0-|4) -- +(\ux,0) node [point] (bx2) {};
	\draw[-,TwoMarks] (bx2)-- (7|-0);
	\draw[-] (3|-0)-- (bx2);
	\draw[-] (7|-0)-- (s);
	
	\end{tikzpicture}
	\vspace{2.2em}
}\hfill
	\subcaptionbox{$`m^*$ applied to the proof of~\eqref{eq:greedy:proof}.\label{fig:greedy:proof}}{
	\def\ux{3em}
	\def\uy{2.5em}
	\begin{tikzpicture}[>=latex]
	\scriptsize
	\node [point] (0) at (0,0) {};
	
	\foreach \j/\p/\w/\wl/\sl in {%
		1/0/6/{$w_{0}=\abs {\mcP}$}/{$s_1=0$},%
		2/1/5/{$w_{e_1}$}/{$s_2=e_1$},%
		3/2.2/4/{$w_{e_j}$}/{$\begin{aligned}&s_{j+1}\\[-.5em] &\kern.5em= e_j\end{aligned}$},%
		4/3.2/3/{$w_{e_{j+1}}$}/{$\begin{aligned}&s_{j+2}\\[-.5em] &\kern.5em= e_{j+1}\end{aligned}$},%
		5/4.4/2/{$w_{e_{\abs{E}}}$}/{$\begin{aligned}&s_{\abs{E}+1}\\[-.5em] &\kern.5em= e_{\abs{E}}\end{aligned}$},%
		6/5.4/1/{}/{$s_{\abs {E}+2}$},%
		7/6.6/1/{$1$}/{$s_{\abs {E}+\abs {V}+1}$}%
	}
	{
		\draw[bar] (\p*\ux,0) -- +(0,\w*\uy) node [point] (\j) {} -- +(1*\ux,\w*\uy)  -- +(1*\ux,0) -- cycle;
		\path (\j)  -- (0|-\j) node (w\j) [htag,label={[label distance=0.1*\u]180:\wl}] {};
		\draw[aline] (\j) -- (w\j.center);
		\path (\j|-0) to node [vtag,label={[label distance=0.4*\u,rotate=-45]0:\sl}] {} +(1*\ux,0);
	};
	
	\foreach \j/\y/\lb in {%
		1/1/$S_1=\Set{0}$%
		,3/3/$S_{j+1}=\Set{0}\cup \Set{e_{j'}\mid j'\leq j}$%
		,5/-1/$S_{\abs{E}+1}=\Set{0}\cup E$%
		,7/-1/$S_{\abs{V}+\abs{E}+1}=S$%
	}
	\draw[decorate, decoration={brace, amplitude=3pt,mirror,raise=0.1*\u}] (\j) +(\ux,0)  node (S\j) {}  -- (0|-S\j) node [midway,above,yshift=0.2*\u,xshift=\y*\u] {\contour{white}{\lb}}; 
	
	\draw[decorate, decoration={brace, amplitude=3pt,mirror,raise=0.4*\u}] (7|-0) +(\ux,0)  node (S7) {}  -- (6|-S7) node [midway,above,yshift=0.6*\u] {$V$};

	\foreach \j/\lb in {%
		1/{$`m^*(S_1)=\abs{\mcP}-w_{e_1}$}%
		,3/$`m^*(S_{j+1})=w_{e_{j}}-w_{e_{j+1}}$%
		,5/$`m^*(S_{\abs{E}+1})=w_{e_{\abs{E}}}-1$%
	}
	\draw[|<->|] (\j) +(1.1*\ux,0) -- +(1.1*\ux,-\uy) node [midway,right,xshift=0.2*\u] {\lb} ; 
	
	\draw[|<->|] (7) +(1.1*\ux,0) -- +(1.1*\ux,-\uy) node [midway,right,xshift=0.2*\u] {$`m^*(S)=1$} ; 
	
	\path (0) to node  {}  (8*\ux,0) node (s) [point,label={[label distance=0.1*\u]right:$s\in S$}] {};
	\path (0) -- (0,7*\uy) node (w) [point,label={[label distance=0.1*\u]above:$w_s$}] {};
	
	\draw[-] (0) --(5-|0);
	\draw[-,TwoMarks] (5-|0)-- (4-|0);
	\draw[-] (4-|0) --(3-|0);
	\draw[-,TwoMarks] (3-|0)-- (2-|0);
	\draw[->] (2-|0)-- (w);
	
	\draw[-] (0) --(2|-0);
	\draw[-,TwoMarks] (2|-0)-- (4|-0);
	\draw[-,TwoMarks] (4|-0)-- (6|-0);
	\draw[-,TwoMarks] (6|-0)-- (S7|-0);
	\draw[-] (S7|-0)-- (s);
	
	\end{tikzpicture}
}
\end{center}
	\caption{Illustration of Edmonds' greedy algorithm in Lemma~\ref{pro:greedy}.}
	\label{fig:greedy}
\end{figure*}

To prove Theorem~\ref{thm:EP}, we will make use of Edmonds' greedy algorithm in combinatorial optimization~\cite[Theorem~44.3]{schrijver02}. A set function $f:2^S\to `R$ with a finite ground set $S$ is said to be \emph{submodular} iff for all $B_1,B_2\subseteq S$,
\begin{align}
	f(B_1)+f(B_2) \geq f(B_1\cap B_2)+f(B_1\cup B_2).\label{eq:submodular}
\end{align}
$f$ is said to be \emph{supermodular} if $-f$ is submodular. If $f$ is both submodular and supermodular, it is said to be \emph{modular}. $f$ is said to be \emph{normalized} if $f(`0)=0$. The entropy function $B\to H(\RZ_B)$, for instance, is a well-known normalized submodular function~\cite{fujishige78,yeung08}. Edmonds' greedy algorithm states that:

\begin{Proposition}[\mbox{\cite[Theorem~44.3]{schrijver02}}]
	\label{pro:greedy}
	For any normalized submodular function $f:2^S\to `R$ with a finite ground set $S$, and any non-negative weight vector $w_S:=(w_s\mid s\in S)\in `R_+^S$, 
	consider the linear program
	\begin{subequations}
		\label{eq:sfo}
		\begin{align}
			&\min_{`m} \sum_{B\subseteq S} `m(B)f(B) \label{eq:sfo:obj}
		\end{align}
	such that $`m:2^S\to `R_+$ is a non-negative set function satisfying
		\begin{align}
			&\sum_{B\subseteq S \colon s\in B }`m(B)=w_s, \kern1em\forall s\in S.  \label{eq:sfo:cons}
		\end{align}
	\end{subequations}
	Then, the optimal solution $`m^*$ to the above problem is given as follows:
	\begin{enumerate}
		\item\label{enum:greedy:1} Enumerate $S$ as $\Set{s_1,...,s_k}$ (with $k:=\abs {S}$) such that 
		$$w_{s_1}\geq \dots \geq w_{s_k}.$$
		\item\label{enum:greedy:2} With $S_j:=\Set{s_{j'}\mid 1\leq j'\leq j}$ for $1\leq j\leq k$, set
		\begin{subequations}
			\label{eq:`m*}
		\begin{align}
			`m^*(S_j)&:=w_{s_j}-w_{s_{j+1}}\kern1em \text {for $1\leq j< k$}\\
			`m^*(S_k)&=`m^*(S):=w_{s_k}
		\end{align}
		\end{subequations}
		and $`m^*(B)=0$ otherwise, i.e., if $B\neq S_j$ for $1\leq j\leq k$.
	\end{enumerate}
	It follows that, if $f$ is modular, the summation in \eqref{eq:sfo:obj} is constant for all feasible $`m$ satisfying \eqref{eq:sfo:cons}.\footnote{This is because $-f$ is submodular and so the same $`m^*$ defined in \eqref{eq:`m*} both minimizes and maximizes the sum in \eqref{eq:sfo:obj}, the value of which must therefore be a constant.}
\end{Proposition}
The algorithm is illustrated in \figref{fig:greedy:general}, which is a plot of $w_s$ against $s\in S$. In particular, the horizontal axis enumerates the elements $S$ in a descending order of their weights $w$ as desired by the greedy algorithm in Step~\ref{enum:greedy:1}. The set of first $j$ elements form the set $S_j$, and the $`m^*(S_j)$ is the drop in height from the $j$-th bar to the $(j+1)$-th bar, with the exception that $`m^*(S_k)$ (or equivalently $`m^*(S)$) is the height of the last bar. 

The proof is by a lamination procedure that can turn any $`m$ to $`m^*$ gradually without increasing the sum in \eqref{eq:sfo:obj} or violating \eqref{eq:sfo:cons}: 
\begin{lbox}
	\noindent\underline{Lamination:}
	For every $B_1,B_2\in \op{supp}(`m)$ such that $B_1$ crosses $B_2$ in the sense that 
	\begin{align*}
		\Set{B_1,B_2}\neq \Set{B_1\cap B_2,B_1\cup B_2},
	\end{align*}
	reduce $`m(B_1)$ and $`m(B_2)$ by $`d$ and increase $`m(B_1\cap B_2)$ and $`m(B_1\cup B_2)$ by $`d$, where
	\begin{align*}
		`d:=\min\Set{`m(B_1),`m(B_2)}\geq 0,
	\end{align*}
	where the non-negativity is by the assumption that $`m$ is non-negative.
	Doing so reduces $\sum_{B\subseteq S} `m(B) f(B)$ by
	\begin{align*}
		`d[f(B_1)+f(B_2)-f(B_1\cap B_2)-f(B_1\cup B_2)]\geq 0,
	\end{align*}
	where the non-negativity is by the submodularity~\eqref{eq:submodular} of $f$.
\end{lbox} 
The procedure turns the support of $`m$ to that of $`m^*$, namely $\Set{S_j\mid 1\leq j\leq k}$, which forms a laminar family (or more specifically, a chain).

Proposition~\ref{pro:greedy} implies an interesting inequality below, which will be frequently used later. 
\begin{Corollary}
\label{cor:lamination}
Consider a finite set $N$ with $0\notin N$ without lose of generality, and a random vector $\RY_N:=(\RY_i\mid i\in N)$ where $\RY_i$'s are mutually independent. Let $\RY_0$ be another random variable which need not be independent of $\RY_N$. For any non-negative set function $`m:2^N\to `R_+$, we have
\begin{align}
\label{eq:sfo:h}
\sum_{B\subseteq N}`m(B)H(\RY_0|\RY_B)\geq \sum_{B\subseteq S}`m^*(B)H(\RY_0|\RY_{B`/\Set{0}})
\end{align}
where $S=\Set{0}\cup N$ and $`m^*(B)$ is obtained by laminating $`m$ as in~\eqref{eq:`m*}. 
\end{Corollary}
\begin{Proof}
Note that $H(\RY_0|\RY_B)=H(\RY_0,\RY_B)-H(\RY_B)$. Let $S=\Set{0}\cup N$ and $		f(B):=H(\RY_B)$  for $B\subseteq S$. It follows that $f$ is normalized and submodular as it is an entropy function of $\RY_S$~\cite{fujishige78}. Then,
\begin{align*}
&\sum_{B\subseteq N}`m(B)H(\RY_0|\RY_B)\\
&=\sum_{B\subseteq N}`m(B)H(\RY_0,\RY_B)-\sum_{B\subseteq N}`m(B)H(\RY_B)\\
&\geq\sum_{B\subseteq S}`m^*(B)H(\RY_0,\RY_B)-\sum_{B\subseteq N}`m(B)H(\RY_B)\\
&=\sum_{B\subseteq S}`m^*(B)H(\RY_0,\RY_B)-\sum_{B\subseteq S}`m^*(B)H(\RY_{B`/\Set{0}}),
\end{align*}
where the first inequality follows from Proposition~\ref{pro:greedy}; the last equality also follows from Proposition~\ref{pro:greedy} since the function $f(B):=H(\RY_B)$ for $B\subseteq N$ is modular, by the assumption that $\RY_i$'s among $N$ are mutually independent. This completes the proof.
\end{Proof}

\begin{Proof}[Theorem~\ref{thm:EP}]
	For any $\mcP\in \Pi'(V)$, by \eqref{eq:recover} and Fano's inequality, 
	\begin{align}
 	 	n`d_n &\geq \sum_{C\in\mcP}  H(\RK|\RF,\tRZ_{C}) \notag\\
		&= \underbrace{\sum_{C\in\mcP} H(\RK,\RF|\tRZ_{C})}_{`(1)}
		-\underbrace{\sum_{C\in\mcP} \kern-.2em H(\RF|\tRZ_{C})}_{`(2)} \label{eq:LB:hyp:123}
	\end{align}
	for some $`d_n\to 0$ as $n\to `8$, where the last equality is by the chain rule expansion. We will bound $`(1)$ and $`(2)$ to obtain the desired lower bound~\eqref{eq:EP}.
	
	$`(2)$ can be bounded by the usual technique (cf.\ \cite[Lemma~B.1]{csiszar08}):
	\begin{align}
	`(2) &\utag{a}=\sum_{C\in\mcP} \sum_{t=1}^{\ell}\sum_{i\in V}H(\RF_{it}|\tRF_{it},\tRZ_{C}) \notag\\
	&\utag{b}\leq \sum_{C\in\mcP}\sum_{t=1}^{\ell}\sum_{i \in V`/C}H(\RF_{it}|\tRF_{it})\notag\\ 
	&\utag{c}=\sum_{t=1}^{\ell} \sum_{i\in V}\sum_{C\in\mcP\colon i\not\in C} H(\RF_{it}|\tRF_{it})\notag\\ 
	&\utag{d}=(\abs{\mcP}-1)\sum_{t=1}^{\ell}\sum_{i\in V} H(\RF_{it}|\tRF_{it}) \notag\\ 
	&\utag{e}=(\abs{\mcP}-1)H(\RF)\label{eq:se_expansion2}
\end{align}
\begin{compactitem}
	\item where \uref{a} follows from the chain rule expansion on $\RF$~\eqref{eq:discussion};
	\item \uref{b} is because
	\begin{align*}
		H(\RF_{it}|\tRF_{it},\tRZ_{C})
		\begin{cases}
			=0  & \text {if $i \in C$ by \eqref{eq:Fit},}\\
			\leq H(\RF_{it}|\tRF_{it}) & \text{otherwise};
		\end{cases}
	\end{align*}
	\item \uref{c} is obtained by interchanging sums;
	\item  \uref{d} is because the summand on r.h.s.\ of \uref{c} is constant with respect to $C$, and so the inner summation gives a multiplicative factor of $\abs {\mcP}-1$.
	\item \uref{e} follows again from the chain rule expansion on $\RF$~\eqref{eq:discussion}.
\end{compactitem}
 
 	Next, we will bound $`(1)$ using \eqref{eq:sfo:h} in Corollary~\ref{pro:greedy}. 	For notational simplicity, define 
 	\begin{align*}
 		E_i&:=\{e\mid i\in `x(e)\} &\kern1em &\text {for }i\in V\\
 		E_C&:=\bigcup_{i\in C} E_i && \text {for }C\subseteq V,
 	\end{align*}
 	which denote the collection of edges incident on node~$i\in V$ and nodes in $C\subseteq V$ respectively.
 	Let $N=V\cup E$ and $S=\{0\}\cup N$, where we assume $0\not\in V\cup E$ without loss of generality. Define $\RY_{S}$ with
 	\begin{subequations}
 		\label{eq:Y}
	\begin{alignat}{2}
		\RY_0&=(\RF,\RK) \\
		\RY_i&=\RU_i &\kern1em& \text{for }i\in V \label{eq:Yi}\\ 
		\RY_e&=\RX_e^n && \text{for }e\in E. \label{eq:Ye}
	\end{alignat}
	\end{subequations}
	It follows that $\RY_s$ for $s\in N$ defined in~\eqref{eq:Yi} and~\eqref{eq:Ye} are mutually independent because of \eqref{eq:U} and the independence of the edge variables.
	Note that $\tRZ_C=(\RU_C,\RZ^n_{C})=(\RU_C,\RX^n_{E_C})$, where the first equality is by~\eqref{eq:tRZ}, and the second equality is by \eqref{eq:Xe}. Hence, we can rewrite $`(1)$ as the sum $\sum_{B\subseteq N}`m(B)H(\RY_0|\RY_B)$ in \eqref{eq:sfo:h} with
	\begin{align*}
		`m(B)&:=
		\begin{cases} 
			1, &B=C\cup E_C, C\in \mcP \\
			0, & \text{otherwise}.
		\end{cases}
	\end{align*} 
	Then, \eqref{eq:sfo:cons} in~Proposition~\ref{pro:greedy} holds with the non-negative weights defined as
	\begin{subequations}
		\label{eq:greedy:proof:w}
	\begin{alignat}{2}
		w_0&:=\sum_{B\subseteq N}`m(B) \notag\\
		&=\sum_{C\in \mcP} `m(C\cup E_C)=\abs{\mcP},  \\
		w_i&:=\sum_{B\subseteq N:i\in B}`m(B)&\kern1em& \text{for $i\in V$}\notag\\
		&=\sum_{C\in\mcP:i\in C} `m(C\cup E_C) =1 \\
		w_e&:=\sum_{B\subseteq N:e\in B}`m(B)&&\text{for $e\in E$}\notag\\
		&=\sum_{C\in\mcP:e\in E_C} `m(C\cup E_C)\notag \\ 
		&=|\Set{C\in \mcP\mid C\cap `x(e)\neq `0}|.\label{eq:w_e}
	\end{alignat}
	\end{subequations}
	As an example, for the triangle PIN $\RZ_{\Set{1,2,3}}$ defined in \eqref{eq:triangle} and illustrated in \figref{fig:triangle}, and the partition $\mcP:=\Set {\Set {1},\Set {2},\Set {3}}$ into singletons, 
	\begin{align*}
		w_0&=\abs {\mcP}=3\\
		w_1&=w_2=w_3=1\\
		w_{\rm{a}}&=w_{\rm{b}}=w_{\rm{c}}=2,
	\end{align*}
	as $w_e$ in \eqref{eq:w_e} reduces to the number of incident nodes of edge $e$ for singleton partition.
	 
	It follows that
	\begin{align*}
		w_0 =\abs {\mcP}\geq w_e \geq 1= w_i  \kern1em \forall e\in E, i\in V.
	\end{align*}
	Enumerate $E$ as $\Set{e_1,\dots,e_{\abs{E}}}$ such that
	\begin{align}
		w_{e_1} \geq w_{e_2} \geq \dots \geq w_{e_{\abs {E}}}.\label{eq:greedy:proof:worder}
	\end{align}
	Then, the desired ordering in Step~\ref{enum:greedy:1} of the greedy algorithm in Proposition~\ref{pro:greedy} satisfies
	\begin{subequations}
		\label{eq:greedy:proof:s}
		\begin{align}
			s_1&=0\\
			\Set{s_2,\dots, s_{\abs{E}+1}}&=\Set{e_1,\dots,e_{\abs{E}}}\\
			\Set{s_{\abs{E}+2},\dots, s_{\abs{E}+\abs{V}+1}}&=V
		\end{align}
	\end{subequations}
	and so $`m^*$ defined in~\eqref{eq:`m*} can be evaluated as shown in \figref{fig:greedy:proof}, with possibly non-zero values at
	\begin{alignat*}{2}
		S_1&=\Set{s_1}=\Set{0}\\
		S_{j+1} &= \Set {0}\cup \Set{e_{j'}\mid 1\leq j'\leq j} &\kern1em & \text {for }1\leq j\leq \abs {E}\\
		S_k &= S = \Set{0}\cup E\cup V.
	\end{alignat*}
	By~\eqref{eq:sfo:h} in Corollary~\ref{cor:lamination}, we can lower bound $`(1)$ with $\sum_{B\subseteq S} `m^*(B)H(\RY_0|\RY_{B`/\Set{0}})$, which simplifies to
	\begin{align}
		\label{eq:greedy:proof}
		\begin{split}
	\kern-0.5em`(1)&\geq \overbrace{(\abs {\mcP}-w_{e_1})}^{`m^*(S_1)} H(\kern-1em\overbrace{\RF,\RK}^{\RY_0|\RY_{S_1`/\Set{0}}}\kern-1em)\\
	&\kern1em+\sum_{j=1}^{\abs{E}-1}
	\overbrace{`1(w_{e_j}-w_{e_{j+1}}`2)}^{`m^*(S_{j+1})} H(\overbrace{\RF,\RK|\RX_{\Set{e_{j'}\mid 1\leq j'\leq j}}^n}^{\RY_0|\RY_{S_{j+1}`/\Set{0}}})\\
	&\kern1em+
	\overbrace{(w_{e_{\abs{E}}}-1)}^{`m^*(S_{\abs {E}+1})}
	H(\kern-1em\overbrace{\RF,\RK|\RX_E^n}^{\RY_0|\RY_{S_{\abs {E}+1}`/\Set{0}}}\kern-1em)\\
	&\kern1em+ H(\overbrace{\RF,\RK|\RX_{E}^n,\RU_{V}}^{\RY_0|\RY_{S`/\Set{0}}}).
	\end{split}
	\end{align}
	Using the triangle PIN and singleton partition again as an example, we have
	\begin{align*}
		`m^*(\Set {0})=`m^*(\Set {0,\rm{a},\rm{b},\rm{c}}) = `m^*(\Set {0,\rm{a},\rm{b},\rm{c},1,2,3})=1
	\end{align*}
	so that the above inequality evaluates to
	\begin{align*}
		&H(\RF,\RK|\tRZ_1)+H(\RF,\RK|\tRZ_2)+H(\RF,\RK|\tRZ_3)\\
		&\kern.2em \geq H(\RF,\RK)+H(\RF,\RK|\RX_{\Set{\rm{a},\rm{b},\rm{c}}})
		+H(\RF,\RK|\RU_{\Set{1,2,3}},\RX_{\Set{\rm{a},\rm{b},\rm{c}}}).
	\end{align*}	
By the fact that entropy and $\mu^*$ are non-negative set functions, 
it follows that
	\begin{align*}
		`(1)&\geq (\abs {\mcP}-w_{e_1})H(\RF,\RK) \\
		&\utag{f}= (\abs {\mcP}-1)`1[1-`a(\mcP)`2] H(\RF,\RK)\\
		&\utag{g}= (\abs {\mcP}-1) `1[1-`a(\mcP)`2] `1[H(\RF)+H(\RK)-n`d'_n`2]
	\end{align*}
	for some $`d'_n\to 0$ as $n\to `8$, where
	\begin{compactitem}
		\item	\uref{f} is because by \eqref{eq:greedy:proof:worder} and \eqref{eq:w_e},
		\begin{align*}
			w_{e_1} := \max_{e\in E} w_e &= \max_{e\in E} \abs {\Set {C\in \mcP\mid C\cap `x(e)\neq `0}}\\
			&= (\abs {\mcP}-1)`a(\mcP)+1 \kern1em  \text {by \eqref{eq:`a}.}\\
			\abs {\mcP}-w_{e_1}&= (\abs {\mcP}-1)`1[1-`a(\mcP)`2] 
		\end{align*}
		\item \uref{g} is by the secrecy constraint~\eqref{eq:secrecy}.
	\end{compactitem}
	Applying the above inequality and \eqref{eq:se_expansion2} to \eqref{eq:LB:hyp:123} and simplifying, we have
	\begin{align*}
		`a(\mcP)\frac{H(\RF)}n 
		\geq `1[1-`a(\mcP)`2] `1[\frac{H(\RK)}n -`d'_n `2]- \frac{`d_n}{\abs {\mcP}-1},
	\end{align*}
which implies \eqref{eq:EP} because  
\begin{align*}
\sup \liminf_{n\to\infty}\frac {H(\RK)}{n}\geq \sup \liminf_{n\to\infty}\frac{1}{n}\log\abs{K}=\CS(R) 
\end{align*}
by the secrecy constraint~\eqref{eq:secrecy} and the definition of secrecy capacity~\eqref{eq:CSR},
\begin{align*}
\limsup_{n\to\infty}\frac {H(\RF)} {n} \leq \limsup_{n\to\infty}\frac{1}{n}\log\abs{F}\leq R
\end{align*}
by~\eqref{eq:rate}, and $`d_n\to 0$, $`d'_n\to 0$ as $n\to `8$. 	
\end{Proof}

\section{Proof of VP bound in Theorem~\ref{thm:VP}}
\label{sec:proof:VP}

Consider any $u_V\in`R_+^V$. By the recoverability constraint~\eqref{eq:recover}, we have
\begin{align*}
n`d_n&\geq \sum_{i\in V} u_i H(\RK|\RF,\tRZ_i)\\
	 &=\underbrace{\sum_{i\in V}u_i H(\RK,\RF|\tRZ_i)}_{`(1)}-\underbrace{\sum_{i\in V}u_{i} H(\RF|\tRZ_i)}_{`(2)}
\end{align*}
We will bound $`(1)$ and  $`(2)$ as follows.
\begin{align*}
`(1)&\geq `1[\sum_{i\in V}u_i-\max_{e\in E}\sum_{i\in V:i\in `x(e)}u_{i}`2]H(\RK,\RF)\\
     &=`1[u(V)-\max_{e\in E}u(`x(e))`2][H(\RK|\RF)+H(\RF)],
\end{align*}
where the inequality is by applying Corollary~\ref{cor:lamination} and keeping only the first term, i.e., $`m^*(S_1)H(\RY_0|\RY_{S_1`/\Set{0}})$. Next, by the fact that conditioning does not increase entropy,
\begin{align*}
`(2)\leq \sum_{i\in V}u_{i}H(\RF)=u(V)H(\RF).
\end{align*}
Now, with
\begin{align*}
\label{eq:varphi}
\varphi(u_V):=\max_{e\in E}\frac{u(`x(e))}{u(V)},
\end{align*}
the bounds on $`(1)$ and $`(2)$ above give
\begin{align*}
`1[1-\varphi(u_V)`2]\frac{H(\RK|\RF)}{n}\leq\varphi(u_V)\frac{H(\RF)}{n}+\frac{`d_n}{u(V)},
\end{align*}
which gives the following bound on $\CS(R)$:
\begin{align*}
\label{eq:VP:varphi}
[1-\varphi(u_V)]\CS(R)\leq \varphi(u_V)R.
\end{align*}

Finally, we will show that the VP bound follows from the above bound by choosing the optimal $u_V$ that minimizes $\varphi(u_V)$ as follows:
\begin{align}
\label{eq:opti:u_V}
\begin{split}
\min_{u_V\in `R_+^V}\varphi(u_V)&=\min_{u_V\in `R_+^V}\max_{e\in E}\frac{u( `x(e))}{u(V)}\\
					    &=\min_{u_V\in `R_+^V: u(\xi(e))\leq 1, \forall e\in E}\frac{1}{u(V)}\\
					    &=\frac{1}{`t}
\end{split}
\end{align}
where the last equality is by~\eqref{eq:`t}. This completes the proof.
As a comparison to the proof of the EP bound in Appendix~\ref{sec:proof:EP}, we did not invoke the inequality~\cite[Lemma~B.1]{csiszar08} for interactive public discussion.

\section{Proof of lamination bound in Theorem~\ref{thm:lb}}
\label{sec:proof:lb}
Similar to the previous proofs, by the recoverability constraint~\eqref{eq:recover}, we have
\begin{align*}
n`d_n&\geq \sum_{B}\pi(B)H(\RK|\RF,\tRZ_{V`/B})\\
         &= \underbrace{\sum_{B}\pi(B)H(\RK,\RF|\tRZ_{V`/B})}_{`(1)}-\underbrace{\sum_{B}\pi(B)H(\RF|\tRZ_{V`/B})}_{`(2)}
\end{align*} 
Again, we will bound $`(1)$ and $`(2)$ separately as follows. 
\begin{align*}
`(1)&=\sum_{B}\pi(B)H(\RK,\RF|\RU_{V`/B},\RX^n_{E_{V`/B}})\\
     &\geq `1[\sum_{B}\pi(B)-\max_{e\in E}\sum_{B: \xi(e)`/B\neq`0}\pi(B)`2]H(\RK,\RF)\\
     &=\underbrace{`1[\min_{e\in E}\sum_{B:\xi(e)\subseteq B}\pi(B)`2]}_{:=\beta(\pi)}`1[H(\RK|\RF)+H(\RF)`2]
\end{align*}
where the inequality follows by applying Corollary~\ref{cor:lamination} and keeping only the first term, i.e., $`m^*(S_1)H(\RY_0|\RY_{S_1`/\Set{0}})$. On the other hand, by~\eqref{eq:rou}, 
\begin{align*}
\kern0.2em`(2)&=\kern-0.3em \sum_{B}\kern-0.1em`1[\pi(B)\kern-0.1em-\kern-0.1em`r`l(B)`2]\underbrace{H(\RF|\tRZ_{V`/B})}_{\utag{a}\leq H(\RF)}\kern-0.1em+`r\kern-0.1em\underbrace{\sum_{B}`l(B)H(\RF|\tRZ_{V`/B})}_{\utag{b}\leq H(\RF)}\\
     &\leq \underbrace{`1[\sum_{B}`1[\pi(B)-`r`l(B)`2]+`r`2]}_{:=`g(`l,`p,`r)}H(\RF)
\end{align*}
where \uref{a} follows from the fact that conditioning can not increase entropy; \uref{b} follows from~\cite[Lemma~B.1]{csiszar08} for the interactive discussion $\RF$ and~\eqref{eq:FP}. It follows that
\begin{align*}	
`b(`p)\frac {H(\RK|\RF)}{n} \leq`1[`g(`l,`p,`r)-`b(`p)`2]\frac {H(\RF)} {n}+`d_n
\end{align*}
which gives the desired reuslt~\eqref{eq:lb} because $`d_n\to 0$ as $n\to `8$,  
\begin{align*}
\sup \liminf_{n\to\infty}\frac {H(\RK|\RF)}{n}\geq \sup \liminf_{n\to\infty}\frac{1}{n}\log\abs{K}=\CS(R) 
\end{align*}
by the secrecy constraint~\eqref{eq:secrecy} and the definition of secrecy capacity~\eqref{eq:CSR}, and
\begin{align*}
\limsup_{n\to\infty}\frac {H(\RF)} {n} \leq \limsup_{n\to\infty}\frac{1}{n}\log\abs{F}\leq R
\end{align*}
by~\eqref{eq:rate}.

\section{Proof of Corollary~\ref{cor:EP:VP}}
\label{sec:proof:EP:VP}

We will show that the lamination bound reduces to the EP and VP bounds by some particular choice of the parameters $`r,`l,$ and $`p$. 

For any $\mcP\in \Pi'(V)$, let 
\begin{align*}
`r=1, \kern1em `l(B)=`p(B)=\begin{cases}
						\frac{1}{\abs{\mcP}-1}, &(V`/B)\in \mcP \\
						0, &\text{otherwise}.
		          \end{cases}
\end{align*}
It follows that
$$`g(`l,`p,`r)=1,$$
\begin{align*}
\kern.5em`b(`p)\kern-.2em&=\min_{e\in E} \sum_{B\in 2^V`/\Set {`0,V}: `x(e)\subseteq B} \pi(B)\\
	 &=\min_{e\in E}\kern-.2em`1[ \sum_{B\in 2^V`/\Set {`0,V}}\kern-.5em`p(B)-\kern-1em\sum_{B\in 2^V`/\Set {`0,V}: `x(e)\cap(V`/ B)\neq`0} \kern-.5em\pi(B)`2]\\
	 &=\min_{e\in E}`1[\frac{\abs{\mcP}}{\abs{\mcP}-1}-\frac{\abs{\Set{C\in \mcP\mid \xi(e)\cap
        C\neq `0}}}{\abs{\mcP}-1}`2]\\
         &=1-`a(\mcP), 
\end{align*}
where the last equality follows by~\eqref{eq:`a}. Then,~\eqref{eq:lb} in Theorem~\ref{thm:lb} becomes
\begin{align*}
 [1-`a(\mcP)] \CS(R) \leq `a(\mcP) R
\end{align*}
which is precisely the EP bound~\eqref{eq:EP}. 

Let $u_V^*$ be the optimal solution to~\eqref{eq:opti:u_V}. Then, set 
\begin{align*}
`r=0, \kern1em `l(B)=0, \forall B\subseteq V, 
\end{align*}and
\begin{align*}
`p(B)=\begin{cases}
					u^*_i,& B=V`/\Set{i},i\in V\\
					0,& \text{otherwise}. 
				    \end{cases}
\end{align*}
It follows that
\begin{align*}
`g(`l,`p,`r)
	       &=\sum_{B\in 2^V`/\Set {`0,V}}`p(B)=\sum_{i\in V}u^*_i=u^*(V)
\end{align*}
\begin{align*}
`b(`p)&= \min_{e\in E} \sum_{B\in 2^V`/\Set {`0,V}: `x(e)\subseteq B} \pi(B)\\
	 &=\min_{e\in E} \sum_{i \in V: `x(e)\subseteq (V`/\Set{i})} u_i^*\\
	 &=\min_{e\in E} \sum_{i \in V`/`x(e)} u_i^*\\
	 &=u^*(V)-\max_{e\in E} u^*( `x(e))
\end{align*}
Therefore, 
\begin{align*}
\frac{`b(`m)}{`g(`l,`p,`r)}=1-\max_{e\in E}\frac{u^*( `x(e))}{u^*(V)}=1-\frac{1}{`t}
\end{align*}
where the last equality follows from the assumption that $u^*_V$ is optimal solution to~\eqref{eq:opti:u_V}.
Applying the above to~\eqref{eq:lb} in~Theorem~\ref{thm:lb} gives the VP bound~\eqref{eq:VP}. 

\section{Alternative converse proof for Example~\ref{eg:lb} via
  iterative lamination}
\label{sec:challenge}
Similar to the proof of the lamination bound, by the recoverability constraint~\eqref{eq:recover}, 
\begin{align*}
n`d_n&\geq H(\RK|\RF,\tRZ_2)+H(\RK|\RF,\tRZ_3)\\
&\kern1em+2H(\RK|\RF,\tRZ_4)+2H(\RK|\RF,\tRZ_5)
\end{align*}
The r.h.s. can be written as $`(1)-`(2)$ where
\begin{align*}
`(1)&:=H(\RK,\RF|\tRZ_2)+H(\RK,\RF|\tRZ_3)\\ &\kern1em +2H(\RK,\RF|\tRZ_4)+2H(\RK,\RF|\tRZ_5)\\
`(2)&:=H(\RF|\tRZ_2)+H(\RF|\tRZ_3)+2H(\RF|\tRZ_4)+2H(\RF|\tRZ_5)
\end{align*}
Again, we will bound $`(1)$ and $`(2)$ separate. We can bound $`(2)$ easily as follows.
  \begin{align*}
  `(2)\leq H(\RF|\tRZ_2)+H(\RF|\tRZ_4)+4H(\RF)
  \end{align*}
  To bound $`(1)$, we will apply the lamination procedure as before.
\begin{align*}
`(1)&\utag{a}=H(\RK,\RF|\RU_2,\RX_{\Set{\rm{a},\rm{c},\rm{d}}}^n)+H(\RK,\RF|\RU_3,\RX_{\Set{\rm{a},\rm{b}}}^n)\\
&\kern1em+2H(\RK,\RF|\RU_4,\RX_{\Set{\rm{b},\rm{c}}}^n)+2H(\RK,\RF|\RU_5,\RX_{\rm{d}}^n)\\
&\utag{b}\geq (6-3)H(\RK,\RF)+(3-2)H(\RK,\RF|\RX_{\Set{\rm{b},\rm{c},\rm{d}}}^n)\\
    &\kern1em+(2-1)H(\RK,\RF|\RU_{\Set{4,5}},\RX_{\Set{\rm{a},\rm{b},\rm{c},\rm{d}}}^n)\\
  &\kern1em+H(\RK,\RF|\RU_{\Set{2,3,4,5}},\RX_{\Set{\rm{a},\rm{b},\rm{c},\rm{d}}}^n)\\
    &\utag{c}\geq 3H(\RK,\RF)+(3-2)H(\RK,\RF|\RX_{\Set{\rm{b},\rm{c},\rm{d}}}^n)\\
    &\geq 3H(\RK,\RF)+H(\RF|\RX_{\Set{\rm{b},\rm{c},\rm{d}}}^n)
\end{align*}
where \uref{a} follows from~\eqref{eq:tRZ} and the source dependence structure~\eqref{eq:scoop}; \uref{b} is obtained by applying Corollary~\ref{cor:lamination} with the appropriate weights calculated in the following matrix;
\begin{center}
  \small
  \begin{tikzpicture}
    \matrix [matrix of math nodes,column sep=-0.1em] {
      |(E1)| & (\RK,\RF) & \RU_2 & \RU_3 &\RU_4 &\RU_5 &\RX_{\rm{a}}^n &\RX_{\rm{b}}^n &\RX_{\rm{c}}^n  & |(d)| \RX_{\rm{d}}^n\\
      |(C1)| 
      H(\RK,\RF|\RU_2,\RX_{\Set{\rm{a},\rm{c},\rm{d}}}^n)&  1   &  1&  &  &  & 1&  & 1& 1\\
      H(\RK,\RF|\RU_3,\RX_{\Set{\rm{a},\rm{b}}}^n)   & 1  &   &1 &  &  & 1& 1&   &   \\
      |(left)| 2H(\RK,\RF|\RU_4,\RX_{\Set{\rm{b},\rm{c}}}^n) &  2  &   &  & 2&  &   & 2& 2 &   \\
      |(5)| 
      2H(\RK,\RF|\RU_5,\RX_{\rm{d}}^n) &  2&   &  &  &  2&   &  &   &   2\\
      |(sum1)|\text{sum}                  & 6& 1&1&2&2&2&3&3 &|(last)|3\\
    };
    \path (5) to node (5sum) {} (sum1);
    \draw (5sum-|left.west) -- (d.east|-5sum);
  \end{tikzpicture}
\end{center}
\uref{c} follows again from Corollary~\ref{cor:lamination} with the appropriate weights calculated in the following matrix;
\begin{center}
  \small
    \begin{tikzpicture}
      \matrix [matrix of math nodes,column sep=-0.1em] {
        |(E2)| & (\RK,\RF)  &\RX_{\rm{b},\rm{c},\rm{d}}^n & (\RU_{\Set{4,5}},\RX_{\rm{a}}^n) &  |(d)| \RU_{\Set{2,3}}\\
        |(C2)| 
        H(\RK,\RF|\RX_{\Set{\rm{b},\rm{c},\rm{d}}}^n)                                   & 1 & 1&     &       &\\
        H(\RK,\RF|\RU_{\Set{4,5}},\RX_{\Set{\rm{a},\rm{b},\rm{c},\rm{d}}}^n)       & 1 & 1&  1& &        \\
        |(10)|H(\RK,\RF|\RU_{\Set{2,3,4,5}},\RX_{\Set{\rm{a},\rm{b},\rm{c},\rm{d}}}^n) & 1 & 1 & 1 &   1&\\
        |(sum2)|\text{sum}                &  3 &3&2&|(last)| 1 \\
      };
     \path (10) to node (10sum) {} (sum2);
      \draw (10sum-|10.west) -- (d.east|-10sum);
    \end{tikzpicture}
  \end{center}
  We call the above bounding technique \emph{iterative lamination} because, unlike the proof of the lamination bound in Appendix~\ref{sec:proof:lb}, Corollary~\ref{cor:lamination} is applied iteratively.
  
Altogether, we have
\begin{align*}
`(1)-`(2)&\geq 3H(\RK,\RF)+H(\RF|\RX_{\Set{\rm{b},\rm{c},\rm{d}}}^n)\\
&\kern1em-`1[H(\RF|\tRZ_2)+H(\RF|\tRZ_4)+4H(\RF)`2]\\
&=3H(\RK,\RF)-5H(\RF)\\
&\kern1em+\underbrace{H(\RF)+H(\RF|\RX_{\Set{\rm{b},\rm{c},\rm{d}}}^n)}_{`(3)}-`1[H(\RF|\tRZ_2)+H(\RF|\tRZ_4)`2]
\end{align*}
We further bound $`(3)$ as follows:
\begin{align*}
`(3)&\utag{a}\geq H(\RF|\tRZ_{\Set{2,5}})+H(\RF|\tRZ_{\Set{1,3,4}})+H(\RF|\RX_{\Set{\rm{b},\rm{c},\rm{d}}}^n)\\
&\utag{b}=H(\RF|\RU_{\Set{2,5}},\RX_{\Set{\rm{a},\rm{c},\rm{d}}}^n)\\ &\kern1em +H(\RF|\RU_{\Set{1,3,4}},\RX_{\Set{\rm{a},\rm{b},\rm{c}}}^n)+H(\RF|\RX_{\Set{\rm{b},\rm{c},\rm{d}}}^n)\\
&\utag{c}\geq H(\RF|\RU_{\Set{2,5}},\RX_{\Set{\rm{a},\rm{c},\rm{d}}}^n)\\ &\kern1em+H(\RF|\RU_{\Set{1,3,4}},\RX_{\Set{\rm{a},\rm{b},\rm{c},\rm{d}}}^n)+H(\RF|\RX_{\Set{\rm{b},\rm{c}}}^n)\\
&\utag{d}\geq H(\RF|\RU_{\Set{1,2,3,4,5}},\RX_{\Set{\rm{a},\rm{b},\rm{c},\rm{d}}}^n)\\ &\kern1em+H(\RF|\RX_{\Set{\rm{a},\rm{c},\rm{d}}}^n)+H(\RF|\RX_{\Set{\rm{b},\rm{c}}}^n)\\
&\geq H(\RF|\RX_{\Set{\rm{a},\rm{c},\rm{d}}}^n)+H(\RF|\RX_{\Set{\rm{b},\rm{c}}}^n)\\
&\geq H(\RF|\RU_2,\RX_{\Set{\rm{a},\rm{c},\rm{d}}}^n)+H(\RF|\RU_4,\RX_{\Set{\rm{b},\rm{c}}}^n)\\
&\utag{e}=H(\RF|\tRZ_2)+H(\RF|\tRZ_4),
\end{align*}
where \uref{a} is because $H(\RF) \geq H(\RF|\tRZ_{\Set{2,5}})+H(\RF|\tRZ_{\Set{1,3,4}})$ by~\cite[Lemma~B.1]{csiszar08}; \uref{b} and \uref{e} follows from the definition~\eqref{eq:tRZ} of $\tRZ_i$'s and the hypergraphical source~\eqref{eq:scoop}; \uref{c} is obtained by applying Corollary~\ref{cor:lamination} to the last two terms; \uref{d} is obtained by applying Corollary~\ref{cor:lamination} to the first two terms.
Therefore, 
\begin{align*}
`(1)-`(2)&\geq 3H(\RK,\RF)-5H(\RF)= 3H(\RK|\RF)-2H(\RF),
\end{align*}
and so
\begin{align*}
3\frac{H(\RK|\RF)}{n}\leq 2\frac{H(\RF)}{n}+`d_n,
\end{align*}
which gives $\CS(R)\leq \frac{2}{3}R$ as desired.

%
%

\bibliographystyle{IEEEtran}
\bibliography{IEEEabrv,ref}

\end{document}